\documentclass[aps,pra,floatfix,epsf,epsfig,nofootinbib,10pt, linesnumbered]{revtex4-2}
\usepackage{tipa}
\usepackage{dcolumn}
\usepackage[sort&compress]{natbib}
\usepackage{bm}
\usepackage[dvips]{epsfig}
\usepackage{graphicx} 
\usepackage{amsmath} 
\usepackage{amssymb}
\usepackage{amsthm}
\usepackage{epsfig}
\usepackage{subfigure}
\usepackage[ruled,vlined]{algorithm2e}
\usepackage{listings}
\usepackage{float}
\usepackage[dvipsnames]{xcolor}
\usepackage{todonotes} 
\usepackage[]{algorithm2e}

\newcommand{\BEQ}{\begin{equation}}
\newcommand{\EEQ}{\end{equation}}
\newcommand{\BEA}{\begin{eqnarray}}
\newcommand{\EEA}{\end{eqnarray}}

\newcommand{\n}{ n}
\newcommand{\m}{ M}

\newcommand{\halpha}{\hat{\alpha}}

\newcommand{\comment}[1]{}

\newcommand{\reprc}{r({\cal C}({\cal T}, {\cal A}))}

\newcommand{\hreprc}{\hat{r}({\cal C}({\cal T}, {\cal A}))}

\begin{document} 
\title{Optimal alphabet for single text compression }

\author{Armen Allahverdyan$^{1)}$ and Andranik Khachatryan$^{1,2)}$}
\affiliation{$^{1)}$Alikhanyan National Laboratory (Yerevan Physics Institute), Armenia\\
$^{2)}$Envoy Media Group, USA
}


\begin{abstract}
A text written using symbols from a given alphabet can be compressed using the Huffman code, which minimizes the length of the encoded text. It is necessary, however, to employ a text-specific codebook, i.e. the symbol-codeword dictionary, to decode the original text. Thus, the compression performance should be evaluated by the full code length, i.e. the length of the encoded text plus the length of the codebook. We studied several alphabets for compressing texts -- letters, $\n$-grams of letters, syllables, words, and phrases. If only sufficiently short texts are retained, an alphabet of letters or two-grams of letters is optimal.
For the majority of Project Gutenberg texts, the best alphabet (the one that minimizes the full code length) is given by syllables or words, depending on the representation of the codebook. Letter 3 and 4-grams, having on average comparable length to syllables/words, perform noticeably worse than syllables or words. Word 2-grams also are never the best alphabet, on the account of having a very large codebook. We also show that the codebook representation is important -- switching from a naive representation to a compact one significantly improves the matters for alphabets with large number of symbols, most notably the words.

Thus, meaning-expressing elements of the language (syllables or words) provide the best compression alphabet.

\comment{
Here we study the optimal noiseless compression of texts using the Huffman code, where the alphabet of encoding coincides with one of those representations. We show that it is necessary to account for the codebook when compressing a single text. Hence, the total compression comprises of the optimally compressed text -- characterized by the entropy of the alphabet elements, and the codebook which is text-specific and therefore has to be included for noiseless (de)compression. 
}

\end{abstract} 

\maketitle

\section{Introduction}

The information theory resolves the problem of noiseless compression for any text written in a particular alphabet
\cite{shannon1,shannon2,cover,yaglom,bell,witten1999ManagingGBs,huffman_review,kallick,raita}. Each alphabet symbol is encoded using a sequence of bits. The codes have different lengths -- frequently occurring symbols get shorter codes. We are interested in lossless compression, meaning we want to be able to recover, without errors, the original text after encoding and decoding. In this setting the shortest possible code-length is bounded by the entropy of the symbols in the text \cite{cover,yaglom}.
 For practical coding, the Lempel-Ziv family of codes, which require a single pass over the data, are often employed \cite{bell,witten1999ManagingGBs}. Huffman codes \cite{huffman_review,raita} are optimal in terms of the code length, but require at least two passes over text \footnote{We emphasize that the optimality of Huffman's code refers to the case when each alphabet symbol is represented by an integral-length codeword. There are more general coding schemes (e.g. arithmetic coding) that go beyond this limitation. They are not relevant for purposes of this work, because for texts they do not provide serious advantages with respect to the Huffman code \cite{raita}. }. 
 
Nonetheless, this well-known structure of optimal noiseless data compression leaves open an important question: what is the best alphabet for compressing a given text? More specifically, any text can be viewed via different representations, i.e. $n$-grams of letters, syllables, words, and phrases. Each of these representations defines a text-specific alphabet and can be used for compression. Which one is preferable? Shannon provided a partial answer to this question \cite{shannon1,shannon2}. He estimated the entropy rate $\frac{1}{\n}S_\n$ calculated from all letter $\n$-grams\footnote{$\n$-grams are the neighbouring sequences of $\n$ letters of the text. We use 26 English letters. Entropy $S_\n$ is defined in (\ref{entropy}).}. If the text is compressed via the alphabet of $\n$-grams, $\frac{1}{\n}S_\n$ bounds the bit-length of the text that is optimally compressed within the alphabet of those $\n$-grams \cite{cover}. Now $\frac{1}{\n}S_\n$ decreases with $\n$ and saturates for $\n\gtrsim 15$ \cite{shannon2,yaglom,king,cover,dembo}. Hence, $\n$-grams with $\n\simeq 15$ provide the optimal alphabet with respect to the bit-length of the coded text. After Refs.~\cite{shannon1,shannon2} introduced entropy methods in studying texts and languages, a long activity of applying information theory methods to languages followed; see e.g. \cite{jakobson,king,yaglom,bell,hilberg,konto,if2,manning,dembo,cover,ebeling,bialek}. One result of this activity is that human subjects tend to produce lower (better) entropy rate estimates than good algorithms, because they understand the meaning of texts \cite{king,konto,if2}. Another result is that $\frac{1}{\n}S_\n$ provides a versatile tool for studying complex systems (including texts and languages) \cite{yaglom,ebeling,bialek,feldman,gras}.
Additionally, a good amount of work has been done for developing fast and efficient text compression algorithms via words \cite{bell,moffat} that also allow to search texts without decompressing them \cite{boa}. However, these methods do not systematically compare words with other alphabets.
Data compression via syllables was also proposed \cite{lansky} based on the idea that any languages normally have much less syllables than words. Since the codebook was not accounted for, Ref.~\cite{lansky} reported an advantage of words (versus syllables) for the (optimal) Huffman code. A marginal advantage for syllables was found only within a sub-optimal (Lempel-Ziv) coding for morphologically rich (fusional) languages \cite{lansky}. 

For a single text compression,  the results by Shannon and others mentioned above are incomplete. Indeed, we need to also account for the codebook length of the text, i.e. for the $\n$-gram-to-codeword correspondence. The codebook is specific for each text, and without it the noiseless decoding of the compressed text is impossible. For letters ($\n=1$) and an average-size human-written text, the codebook length indeed tends to be irrelevant, as we show below. Moreover, for letters codebooks are nearly universal for a typical and sufficiently long English text, because letters will tend to appear there with stable, text-independent frequencies. As shown below, codebooks lengths will be relevant already for $\n\geq 2$, i.e.  well before the Shannon bound $\n\simeq 15$.

Here we want to determine the optimal alphabet for full code length, i.e. the bit-length of the coded text plus the codebook length. To this end, we analyzed ~$30,000$ books (texts) of Project Gutenberg \cite{gutenberg}, where for each text the compression alphabet ${\cal A}$ is chosen to be $\n$-grams of its letters ($\n=1,2,3..$), or its syllables\footnote{
Appendix \ref{ap_a} discusses syllables and peculiarities of their definition.}, or words\footnote{A useful but not completely exhaustive definition of word is provided by Bloomfield \cite{bloom}: A word is a form which may be uttered alone (with meaning) but cannot be analyzed into parts that may (all of them) be uttered alone (with meaning). For example, {\it books} is a word, since {\it book} can be uttered alone (with meaning), while {\it s} cannot be. In modern English, as well as in some other languages, words are typically shown with a space on either side when written or printed. This is sometimes taken as the definition of word, but note that this need not be the case in other writing system; e.g. Classical Latin and Late Latin frequently did not employ space symbols.}, or pair of words. 
Now for which ${\cal A}$ the full code-length (code-only length + codebook length) will be minimal?

Our main result is the optimal alphabet for texts from Project Gutenberg is provided either by syllables or words, depending on the codebook representation.
A more complete representation favors syllables, while the most compact representation favors words. This result changes if Project Gutenberg texts are filtered with respect to their length: if sufficiently short texts are retained, the optimal alphabet is letters (for shorter texts) or 2-grams of letters (for longer texts). In the language hierarchy|letters, syllables, words, phrases \textit{etc}|syllables is the first level that starts to express meaning, as opposed to letters or phonemes which merely form it \cite{bloom}. The meaning-expression is represented more completely with words: all words are meaningful, while many syllables are not. Note that the relevance of meaning is apparent in our results: the average length of an English syllable is $\simeq 3$ letters, but syllables are noticeably better as an alphabet for compression than both $3$-grams and $4$-grams. We apply these results for discussing compositional {\it versus} combinatorial structure of the language; see section \ref{summary}.

These results hold only once the codebook length is accounted for. 
Otherwise, if the code-only length is taken alone, more concatenated alphabets are better; e.g. words are better than syllables, and letter 4-grams are better than 3-grams. Moreover, we present a compact representation for the codebook and show that it is important. In particular, a naive, non-economical representation biases the results towards favoring shorter alphabets -- letters and letter 2-grams. 

This paper is organized as follows. Next section fixes notations and reviews the standard set-up of noiseless compression; see Table~\ref{tab0} for notations employed. Section \ref{co} discusses the codebook of compression and its representation via bits. Our results on texts from Project Gutenberg  are presented in section \ref{results}. We summarize and discuss relations with literature in the last section. In particular, we discuss our results in the context of emergence of meaning-expressing elements of communication (words and syllables) from meaningless, but meaning-distinguishing elements (letters and phonemes). Appendices \ref{ap_a}, \ref{codebook_letters} and \ref{pseudo} discuss 
technical questions. Appendix \ref{univ} studies uncompressed codebook representations.  
Appendix \ref{when} analyzes theoretically whether more concatenated codes would result in shorter code (only) length. Appendix \ref{kolmo} discusses how our results can apply for bounding Kolmogorov complexity of texts. 

\begin{table}
\caption{\label{tab0}Notations used in the paper (possibly) together with 
equation numbers that define or discuss them.}
\centering
\begin{tabular}{|c||c|} 
\hline
${\cal T} = \langle \tau_j \rangle_{1 \leq j \leq N}$  & text composed of $N$ (generally not distinct) symbols  $\tau_j$. \\ \hline
${\cal A}=\{a_k\}_{k=1}^\m$ & alphabet of distinct symbols $a_k$. \\ \hline
${\cal C}({\cal T}, {\cal A})$ & codebook of the text ${\cal T}$ encoded by via alphabet ${\cal A}$; see (\ref{codebook}).\\ \hline
$enc({\cal T})$ & encoded text ${\cal T}$; see (\ref{wash}).\\ \hline
$\ell[b]$ & bit-length for a binary string $b$; see (\ref{hu}).\\ \hline
$S[f]$ & entropy defined by (\ref{entropy}).\\ \hline
$\gamma(b)$ & self-delimiting code for a binary string $b$; see (\ref{eq:gamma-code}).\\ \hline
$x_2$ & is the number $x$ written in binary. \\ \hline
$\lceil x\rceil$ & smallest positive integer $\geq x$. \\ \hline
$\reprc$ & decodable binary representation of the codebook ${\cal C}(\cal T, \cal A)$, defined via 
(\ref{eq:codebook-in-blocks}, \ref{dodo}). \\
\hline
$letters({\cal T})$ & 
number of letters in text ${\cal T}$. \\ \hline $\eta[{\cal T}]$ & compressibility defined by (\ref{fox}).\\ \hline
$\m_{\rm words}$ & number of distinct words in a text; see (\ref{grund1}, \ref{grund2}).\\ \hline
$N_{\rm words}$ & number of all words in a text; see (\ref{grund1}, \ref{grund2}).\\ \hline
$L=8$ & ASCII representation of codebook symbols letters; see section \ref{godoy} and (\ref{nebed4}). \\ \hline
$L={\rm variable}$ & Huffman representation of codebook symbols letters; see section
\ref{godoy} and (\ref{nebed5}). \\ \hline
\end{tabular}
\end{table}

\section{Noiseless compression of a single text}
\label{noiseless}

A text ${\cal T} = \langle \tau_j \rangle_{1 \leq j \leq N}$ is a sequence of $N$ symbols where each symbol $\tau_j$ is drawn from the alphabet ${\cal A}=\{a_k\}_{k=1}^\m$. In our case ${\cal T}$ will be a text from  Project Gutenberg, while ${\cal A}$ will be its distinct letters (possibly including punctuation marks), distinct $\n$-grams of letters, distinct syllables, distinct words, distinct pairs of words. 

A noiseless code maps each element $a_k$ to a sequence of bits $code(a_k)$ such that any text written in ${\cal A}$ is uniquely decodable \cite{cover}. The simplest uniquely decodable codes are prefix-free codes, where none of $code(a_k)$ is a prefix for $code(a_l)$ ($k\not=l$); e.g. $code(a_1)=0$ and $code(a_2)=01$ are not allowed. Now 
\BEA
\label{codebook}
{\cal C}({\cal T}, {\cal A})=\{a_k\Leftrightarrow code(a_k)\}_{k=1}^\m, 
\EEA
is the codebook of the code. Generally, it depends on both ${\cal T}$ and ${\cal A}$. To encode the text, we encode each symbol in ${\cal T} = \langle \tau_j \rangle_{1 \leq j \leq N}$ and concatenate the codewords:
\BEA
\label{wash}
enc({\cal T}, {\cal C}({\cal T}, {\cal A})) = code({\tau}_1)code(\tau_2) \ldots code(\tau_N).
\EEA
We will write (\ref{wash}) as $enc({\cal T})$ for brevity. Denote with $\ell[b]$ the length of a bit sequence $b$. For each ${\cal T}$, consider the uniquely decodable code that minimizes the bit-length $\ell[enc({\cal T})]$ of $enc({\cal T})$ \cite{shannon1,cover}:
\BEA
\label{hu}
\ell[enc({\cal T})] = \sum_{j=1}^{N}{\ell[code(\tau_j)]} = \sum_{k=1}^\m {m_k \cdot {\ell}[code(a_k)]},
\EEA
where $m_k$ is the number of times $a_k$ appeared in ${\cal T}$, and $\ell[code(a_k)]$ is the bit-length of $code(a_k)$ \cite{cover}. The general idea of minimizing (\ref{hu}) is that more frequent symbols are represented via shorter code-words. Now we employ a known theorem of information theory, which states that the minimum of (\ref{hu}) is achieved via the prefix-free Huffman code \cite{cover}. The minimal (optimal) value of (\ref{hu}) over all uniquely decodable codes is bounded by \cite{cover}:
\BEA
\label{or}
N S[f]<\sum_{k=1}^\m m_k \cdot \ell[code(a_k)] \leq N S[f]+c N, \qquad c<1,\\
S[f]=\sum_{k=1}^{\m} f[a_k]\log_2\left[\frac{1}{f[a_k]}\right],\qquad
f[a_k]=\frac{m_k}{N},\qquad 
\sum_{k=1}^{\m} f[a_k]=1.
\label{entropy}
\EEA
where $f[a_k]$ is the frequency of $a_k$ in ${\cal T}$, $S[f]$ is entropy, and where $c<1$ (for Huffman's code) depends on ${\cal T}$ and ${\cal A}$ and does not have a universal expression, though it holds tangible inequalities \cite{gallager,mans}. Thus, within the family of prefix-free coding methods that achieve (\ref{or}), Huffman's method provides the provably minimal value of $c$. 

Note that for a sufficiently large $\m$, the second term $cN$ in (\ref{or}) tends to be smaller than the first term. To illustrate the emergence of entropy (\ref{entropy}) in (\ref{or}), we can employ the sub-optimal Shannon coding method \cite{shannon1,cover}, where $c=1$, but now the method is straightforward \cite{cover}: 
starting from more probable symbols $a_k$ one chooses $code(a_k)$ among lexicographically first bit-sequences that have the length $\lceil\log_2 ({1}/{f[a_k]}) \rceil$ and maintain the prefix-free feature. From now on we imply the Huffman code, i.e. the minimal value of $c$ in (\ref{or}). Note that we do not employ the Lempel-Ziv family of compression algorithms \cite{cover}, because they are sub-optimal, i.e. they provide a larger value of the average length (\ref{hu}) as compared to Huffman's code. The advantage of this family is that its representatives work faster, but here we are not interested in issues related to runtime speed. 

\section{Codebook}
\label{co}

\subsection{Codebook representation and its length}
\label{subsec:Codebook-CodebookRepresentation}

The optimal code for ${\cal T}$ does necessarily have a text-specific codebook ${\cal C}({\cal T}, {\cal A})$ in (\ref{codebook}), because e.g. the same word can have different frequencies in different texts if ${\cal A}$ amounts to distinct words of ${\cal T}$. Hence ${\cal C}({\cal T}, {\cal A})$ should be used together with the encoded text $enc({\cal T})$ to decode the text. 

We need to represent ${\cal C}({\cal T}, {\cal A})$ via bits and add the bit-length to (\ref{hu}). 
The decoder, upon receiving the code for the codebook and the text, is going to first decode the codebook and then use that codebook to decode the text. Therefore, the codebook encoding should adhere to a certain agreed-upon scheme, so that the decoder can understand it without having a reference to the text itself. Such a representation should also be compact.

For ${\cal A}$ we are employing letter $\n$-grams, syllables, words, and phrases. Therefore, a single entry in the codebook maps a sequence of letters (and possibly special symbols), e.g. a letter $\n$-gram, or a syllable {\it etc.}, to its Huffman codeword; cf.~(\ref{codebook}). To describe a text-agnostic, decodable representation of the codebook, we introduce an auxiliary coding scheme which encodes an arbitrary binary string into a self-delimiting representation.

\subsubsection{Self-delimiting representation}

Given a bit-string $b$, we define a self-delimiting code for $b$, called $\gamma$-code (cf.~\cite{elias,LiVitanyi}), as follows:
\BEA
\label{eq:gamma-code}
\gamma(b) = \underbrace{000...0}_{\ell[\ell[b]_2] \; \rm{zeros}} \; \ell[b]_2 \; b
\EEA
where $\ell[b]$ is the length of the binary code $b$, and $\ell[b]_2$ is the number $\ell[b]$ written in binary. We can also apply $\gamma(\mu)$ to integer numbers $\mu$, by writing $\mu$  in binary:
$\gamma(\mu)=\gamma(\mu_2)$. 
\\
\textbf{Example}:
Let $b={\rm 1001101}$, then $\ell[b] = 7$, $\ell[b]_2 = 7_2 = 111$, and $\ell[\ell[b_2]] = 3$. The $\gamma$-code becomes
\BEA
\label{1star}
\gamma(1001101) = 000 \; 111 \; 1001101 .
\EEA
To obtain $b$ from $\gamma(b)$, we start reading 0-s until we see the first 1. The number of zeros before the first 1 is the length of $\ell(b)_2$, which is 3 in this example. The next 3 bits are the binary representation of $\ell[b]_2$, which is $111$. We now know the length of the $b$ -- it is $111 = 7_2$. The next 7 bits is $b$ itself.
\qed

Given an arbitrary set of binary strings $\cal B$, the set of codes $\gamma({\cal B}) = \{\gamma(b) | b \in {\cal B}\}$ are prefix-free, since we always know where $\gamma(b)$ ends. The length of the $\gamma$-code for $b$ is 
\BEA
\label{2star}
\ell[\gamma(b)] = \ell[b] + 2{\lceil} \log_2{\ell[b]} {\rceil} = \ell[b] + 2 \log_2{\ell[b]} + O(1).
\EEA

\subsubsection{Representation of codebook} 
\label{codo}

Denote the decodable, binary representation of the codebook as $\reprc$. We want to use a representation which is as compact as possible, but is also text-agnostic, i.e. the codebook can be recovered from this representation in a text-independent way. A naive  representation of the codebook would be to list the alphabet symbols followed by their Huffman codes. This representation is uneconomical because we can use the Huffman code to recover the codes even if we do not send them, but send instead their bit-lengths \cite{kallick,raita}. Moreover, we do not need to send individual code-lengths as well, we can send (for all involved Huffman code-lengths) the number of alphabet symbols having the same code-length \cite{kallick,raita}. Hence, instead of sending the number $6$ (say) 10 times, it suffices to send $6$ and $10$. 

Note that the same feature holds for the Shannon sub-optimal code; cf.~the discussion after (\ref{entropy}). Also for this method it suffices to send the number of alphabet symbols having codewords of the same length. For the Shannon method the origin of this feature is simpler and stems from the fact that the codewords are chosen among lexicographically first bit-sequences that have a given length and maintain the prefix-free property \cite{cover}. For the Huffman method this feature needs a special implementation of the method \cite{kallick,raita}.

To represent the codebook, we sort the symbols in $\cal A$ by their Huffman code length. We then group symbols with the same Huffman code length into a "block":
\BEA
\label{eq:codebook-in-blocks}
a_1^{(t)}\ldots a_{k_t}^{(t)} \;\;\; a_1^{(t+1)} \! \ldots a_{k_{t+1}}^{(t+1)}\;\;\; a_1^{(T)}\ldots a_{k_T}^{(T)}
\EEA
where for an integer $z$, $k_z$ alphabet symbols $a_1^{(z)}\ldots a_{k_t}^{(z)}$ have Huffman code-length equal to $z$. The full number of alphabet symbols is then $M=\sum_{z=t}^Tk_t$.

We encode (\ref{eq:codebook-in-blocks}) block-by-block.
For an integer $z$, we encode the block corresponding to $z$ as follows [cf.~(\ref{eq:gamma-code}--\ref{2star})]:
\BEA
\gamma(k_z)\; \gamma(z)\;\;\; \gamma(\alpha(a_1^{(z)})) \; \ldots 
\; \gamma(\alpha(a_{k_z}^{(z)})),
\label{dodo}
\EEA
where $\alpha(a_k)$ is a representation of an alphabet symbol $a_k$. In (\ref{dodo}) we know when the current block ends, and hence each next block can be decoded unambiguously.

\subsubsection{Representation of letters}
\label{godoy}

We focus on two methods for representing $\alpha(a_k)$ via bits. First, we represent each letter of $a_k$ using a bit-sequence of length $L$. This $L$-bit letter-to-bits mapping is shared beforehand between the encoder and the decoder. We take $L=8$, since this is the standard ASCII code for letters. We refer to this as $L=8$ representation. 

Second, we represent each letter of $a_k$ via Huffman code obtained through the standard frequency of that letter, which is estimated from a big dataset of English texts; see Appendix \ref{codebook_letters}. Since the Huffman coding is prefix-free, and the binary code for each letter can be known {\it apriori}, this representation is both compact and text-agnostic. Also, since the letter frequencies are more or less stable for English texts, we can expect that this representation is in fact close to the optimal (not necessarily text-agnostic) compression of alphabet symbols via  letters. As compared to the first representation, this second method of representing $\alpha(a_k)$ has the following drawback: real texts contain more symbols than letters, and such non-letters are to be either taken away during lemmatization, or represented differently. We choose the first option and exclude non-letter symbols from texts.
We refer to this as $L={\rm variable}$ representation. 

Let us emphasize already here that this representation of $\alpha(a_k)$ produces results that are not far from taking $L=5$ instead of $L=8$; see Table \ref{table1}. Now $L=5$ amounts to a lemmatization of texts, where we keep 26 English letters plus a few (not more than 32-26=6) additional text symbols. Appendix \ref{pseudo} discusses a pseudo-code for the codebook encoding. 

\subsubsection{Length of the codebook representation}

Returning to (\ref{dodo}), we note that for obtaining its overall representation, we need to add the number of blocks $T-t$, so that the decoder knows when to stop. This implies an additional overhead of $\ell[\gamma(T-t)]$ bits, and then the length of the codebook representation reads from (\ref{dodo}):
\BEA
\label{nebed1}
\label{eq:codebook-length-block-compr}
{\ell}[\reprc] &=& \ell[\gamma(T-t)]+
\sum_{z=t}^T \left\{\ell[\gamma(z)]+\ell[\gamma(k_z)]\right\}+
\sum_{a \in {\cal A}} \ell[\gamma(\alpha(a))],
\EEA
where we used $\sum_{z=t}^T\sum_{u=1}^{k_z}\ell[\gamma(\alpha(a^{(z)}_u))] = \sum_{a \in {\cal A}} \ell[\gamma(\alpha(a))]$. 
For the first method of representing $\alpha(a_k^{(z)})$ via $L$-grams of letters, we have
\BEA
\label{nebed4}
\sum_{a \in {\cal A}} \ell[\alpha(a)]=
L \times \sum_{a \in {\cal A}}{letters(a)},
\EEA
where $letters(a)$ is the number of letters in $a$.
For the second method of representing letters the term $\sum_{a \in {\cal A}} \ell[\alpha(a)]$ is to replaced by its proper Huffman expression. The latter can be bounded from above via [cf.~(\ref{entropy})] 
\BEA
\sum_{a \in {\cal A}} \ell[\alpha(a)]\lesssim
\sum_{a \in {\cal A}}{letters(a)}\times \sum_{k=1}^{26}\theta_k\left\lceil\log_2\frac{1}{\theta_k} \right\rceil=4.5766\times \sum_{a \in {\cal A}}{letters(a)},
\label{nebed5}
\EEA
where $\sum_{k=1}^{26}$ goes over all letters, and where $\theta_k$ is the frequency of letter $k$. The last equation in (\ref{nebed5}) refers to the frequencies we employed; see Appendix \ref{codebook_letters}.

\subsection{Compressibility}

Recall that the original (uncompressed) text ${\cal T}$ (written with alphabet of $\m$ symbols) can be represented via $N\lceil \log_2 \m\rceil$ bits, assuming that each symbol of ${\cal A}$ is represented via $\lceil \log_2 \m\rceil$ bits. This motivated the definition of redundancy $\ell[enc({\cal T})]/(N\lceil \log_2 \m\rceil)$ that is not larger than 1 after the optimal compression [see (\ref{hu}, \ref{entropy})], because $S[f]\leq \log_2 \m\leq \lceil \log_2 \m\rceil$ \cite{cover}. However, the redundancy does not characterize the degree of text compression. First, even if we agree to represent the alphabet elements by $\lceil \log_2 \m\rceil$ bits, this alphabet is still text specific, e.g. the set of distinct words of ${\cal T}$ is specific for ${\cal T}$. Second, we do need to include the codebook length into the definition of the compression degree.

To define compressibility, we shall assume that ${\cal T}$ is just given in a form that is available for reading in standard computers: its letters are represented via the ASCII code, where each letter is coded via $L$ bits ($L=8$, as discussed above). Hence, initially ${\cal T}$ is represented via $L\times letters({\cal T})$, where $letters({\cal T})$ is the number of letters in ${\cal T}$, if necessary including punctuation marks and special symbols. Then compressibility $\eta[{\cal T}]$ reads
\BEA
\eta[{\cal T}]=\frac{\ell[enc({\cal T})]+\ell[\reprc]}{L\times letters({\cal T})},
\label{fox}
\EEA
where $\ell[\reprc]$ in (\ref{fox}) is defined via the same value of $L$; cf.~(\ref{eq:codebook-length-block-compr}). Now $\eta<1$ means that some compression was achieved; see \cite{we} for related ideas on compressibility. 

\comment{
\BEA
\label{eq:Kolmogorov-complexity-text}
K[{\cal T}]&&\leq \ell[\gamma(enc({\cal T}))] + \ell[\gamma(n)] + \ell[\reprc] + O(1) \\
&& = \ell[\reprc] + X + 2\log_2{X} \nonumber \\
X && = {\cal S}[{\cal T}, {\cal A}] + cN + \log_2(n) + O(1) \nonumber
\EEA
}

\begin{figure}[!h]
\centering
    \subfigure[]{
    \includegraphics[width=8.6cm]{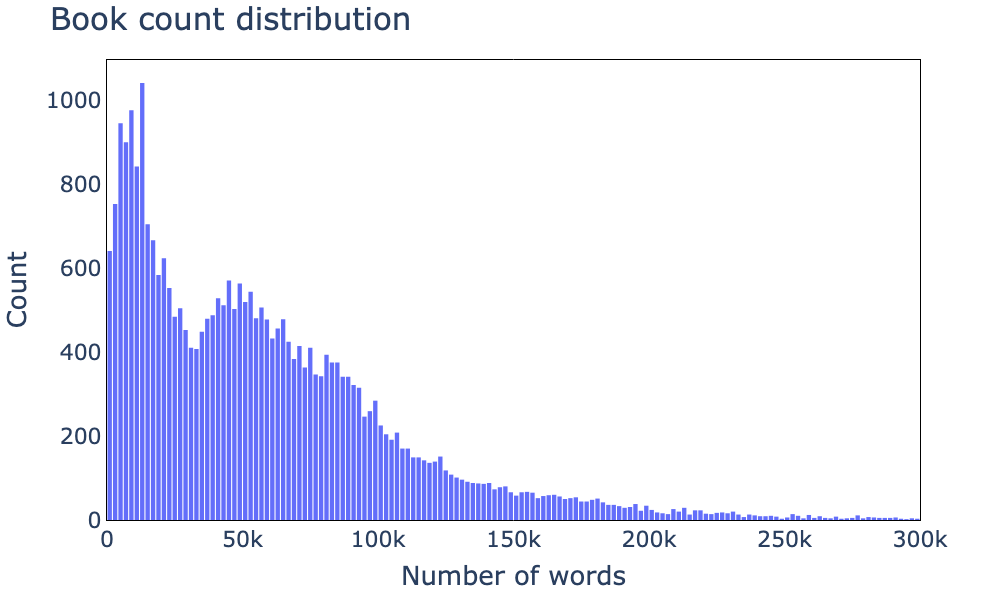}
    \label{fig:word-count-histogram} 
    }
    \subfigure[]{
    \includegraphics[width=8.6cm]{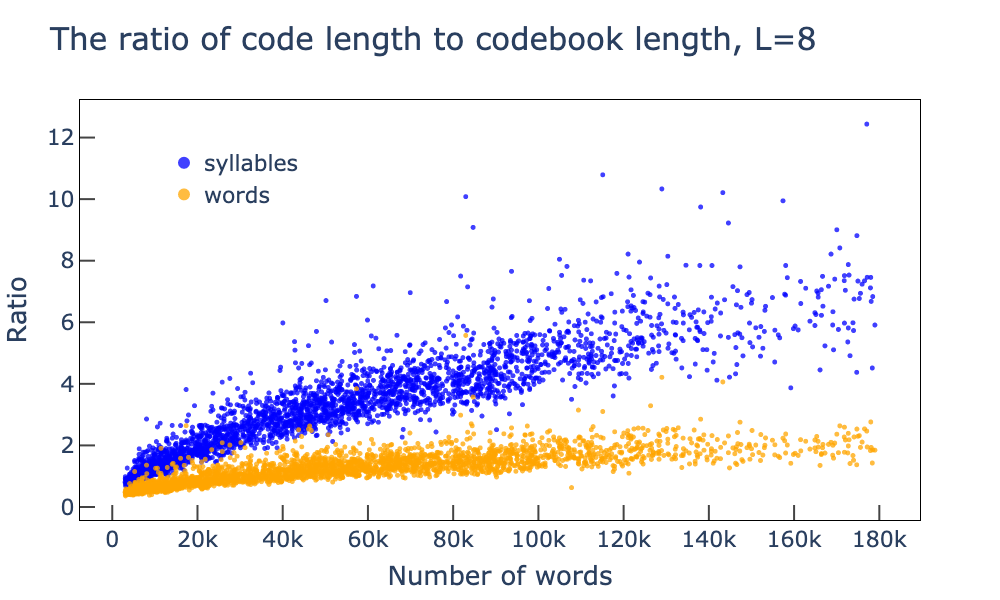}
    \label{fig:codebook-vs-code-size} 
    }
    \caption{(a) Distributions of Project Gutenberg texts over their lengths in words. Here $50{\rm k}=50\times 10^3$. \\
    (b) The ratio $\ell[enc({\cal T})]/\ell[\reprc]$ of the code-only length to the codebook length for alphabets of words (lower set of orange points) and syllables (upper set of blue points) within the $L=8$ (ASCII) representation of codebook; cf.~(\ref{nebed4}). We have $\ell[enc({\cal T})]/\ell[\reprc]\lesssim 2$ and $\ell[enc({\cal T})]/\ell[\reprc]\lesssim 8$ for (most of) words and syllables, respectively. Qualitatively the same picture holds for the $L={\rm variable}$ representation of the codebook. But now $\ell[enc({\cal T})]/\ell[\reprc]\lesssim 4$ and $\ell[enc({\cal T})]/\ell[\reprc]\lesssim 15$ for (resp.) words and syllables. Such a change is natural, since $L={\rm variable}$ provides a more compact codebook representation.
    }
\end{figure}

\begin{figure}[!h]
\centering
    \subfigure[]{
    \includegraphics[width=8.6cm]{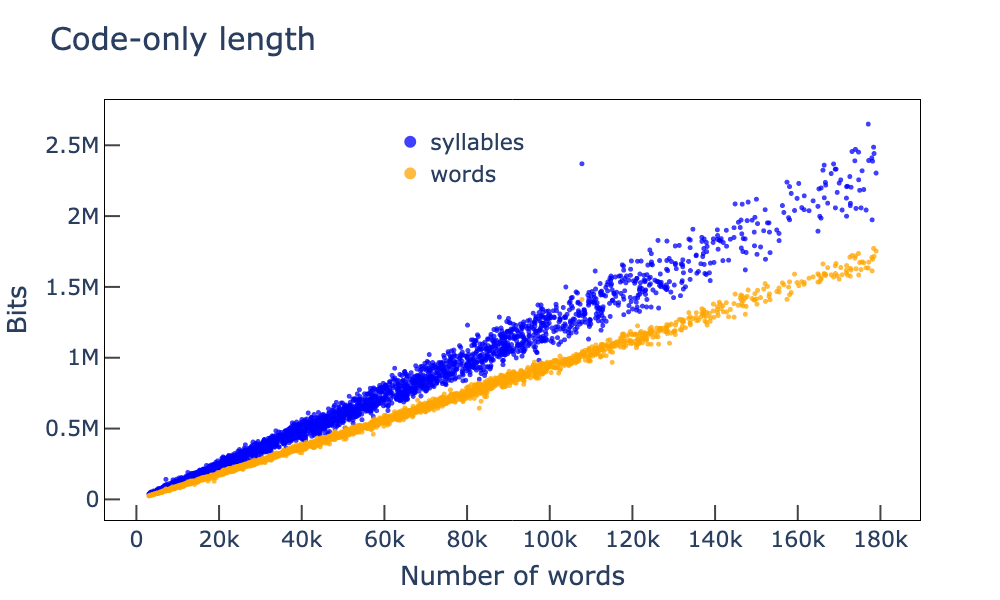}
    \label{fig:code-only-length} 
    }
    \subfigure[]{
    \includegraphics[width=8.6cm]{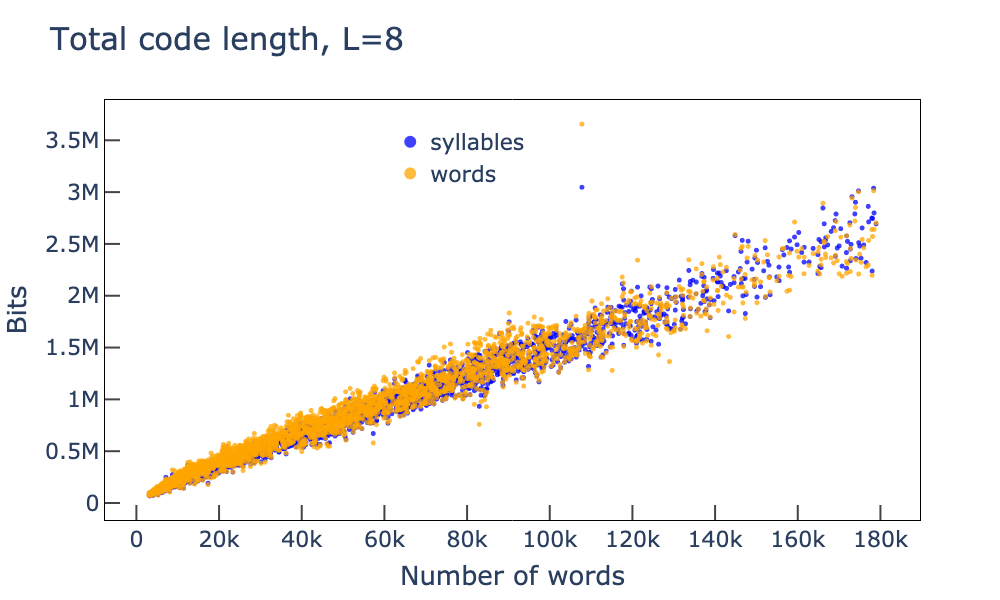}
    \label{fig:total-code-length} 
    }
    \caption{(a) The code-only (bit) length $\ell[enc({\cal T})]$ for Project Gutenberg texts versus the total number of words $N_{\rm words}$ in them. $\ell[enc({\cal T})]$ is the Huffman encoded version of $\ell[{\cal T}]$; cf.~(\ref{hu}--\ref{entropy}). Blue (upper) and orange (lower) points denote (resp.) the alphabet of distinct syllables and words for each text. Here $1{\rm M}=10^6$ and $1{\rm k}=10^3$. As seen from Appendix \ref{when}, words (being a more concatenated alphabet than syllables) provide lower values of $\ell[enc({\cal T})]$.\\
    (b) The same as in (a), but for the total code-length (in bits) $\ell[enc({\cal T})]+\ell[\reprc]$, where the codebook length $\ell[\reprc]$ is calculated according to $L=8$; cf.~(\ref{nebed4}). Now there is no clear-cut relation between syllables and words: the advantage varies from one book to another; see Table~\ref{table1}.
    }
\end{figure}

\section{Compressing texts from Project Gutenberg}
\label{results}

\subsection{Text selection, lemmatization and other technical details }

We studied all books of Project Gutenberg, which amounts to $30304$ English texts. Fig.~\ref{fig:word-count-histogram} shows that the word-count distribution of these texts have two maxima. These texts were ranked according to their length in words. Then we excluded certain anomalous texts; in particular, the shortest 100 and the longest 100 texts were excluded. We ended up with 29708 texts. We randomly selected 10 \% of these texts retaining their original ranks. All figures (besides Fig.~\ref{fig:word-count-histogram}) were constructed out of this sample. Table~\ref{table1} was made out of full 29708 texts. 

For each text we made all letters lowercase, and omitted  numbers and punctuation marks. We do not code the space symbol, i.e. we do not recover the space symbol after decoding. This is a minor issue for two reasons. First, we checked that a decoder knowledgeable in English can recover different words with a negligible error. Second, if we do not want to tolerate even this small error, spaces and punctuation marks can be recovered using an auxiliary code which does not depend on the choice of the encoding alphabet; i.e., omitting it from everywhere should not have a significant effect. Also letter $\n$-grams are defined as a sequence of $\n$ consecutive letters neglecting the space symbol. Singular and plural are considered as different words. For syllabification we employed routine \cite{pyphen}, which is based on the syllabification libraries by Hunspell \cite{hunspell}; see Appendix \ref{ap_a}. 
Note that this Appendix evaluates two widespread syllabification methods with respect to their compression ability and concludes that they are nearly equivalent. 

\comment{For pairs of consecutive words, we add a delimiter between the words during encoding and remove it after decoding. }

Each text is compressed via Huffman's code employing the alphabets of letters, $\n$-grams of letters ($\n=2,3,4$), (distinct) syllables,  words, and pairs of words. Even for the alphabet of letters, the codebook is text-specific, because the same letter can have different frequencies in different texts. Hence, everywhere we account for both the encoded text length (code-only length) and the total code length (code-only plus codebook); see section \ref{co}. The codebook length is important for compression via alphabets of syllables and words; see Fig.~\ref{fig:codebook-vs-code-size}. It is not important for the alphabet of letters, as seen below.

\begin{table}[h!]
\caption{ Comparison between various alphabets with respect of the total code-length for all 29708 books of Project Gutenberg. Here $ \{\rm s< all\}$ means the ratio (percentage) of books for which syllables as an alphabet provide the lowest total code-length compared to other studied alphabets: words (w), letters (lett), 2-grams of letters (2lett). Likewise, $ \{\rm lett2< w\}$ denotes the ratio of cases where the compression via the alphabet of letter 2-grams provides a lower total compression length compared to the alphabet of words. \\
 $L=8$ means that each codebook letter is represented via a fixed string of $8$ ASCII bits; cf.~section \ref{godoy}. $L=5$ is the compressed analogue of $L=8$, where every letter is represented via $5$ bits; hence only 26 letters can be represented plus $2^5-26=6$ additional symbols; $L={\rm variable}$ means the variable-length representation of codebook letters; see section \ref{godoy}. It is seen that alphabets with longer symbols get advantage when using more compact codebook representations; e.g. words win over other alphabets for $L=5$ and $L={\rm variable}$. }
\centering
\begin{tabular}{ |c||c|c|c|c||c|c|c|c| } 
 \hline
{\rm codebook}& $ \{\rm s< all\}$ & $ \{\rm w< all\}$ & $ \{\rm lett2 < all\}$ & $ \{\rm lett< all\}$ & $ \{\rm s< w\}$ 
       & $ \{\rm lett< w\}$ & $ \{\rm lett2< w\}$ & $ \{\rm w< lett2\}$   \\
       \hline
 $L=8$ & 0.497 & 0.175 & 0.212 & 0.116 & 0.810 & 0.375 & 0.477 & 0.523  \\ 
       \hline
 $L=5$ & 0.169 & 0.661 & 0.110 & 0.061 & 0.300 & 0.161 & 0.184 & 0.816  \\ 
       \hline
 $L={\rm variable}$ & 0.061 & 0.816 & 0.071 & 0.052 & 0.130 & 0.109 & 0.116 & 0.884  \\ 
 \hline
\end{tabular}
  \label{table1}
\end{table}

\subsection{Results}

\subsubsection{Codebook length is important: estimates based on the Zipf's law}
\label{zipo}

Fig.~\ref{fig:codebook-vs-code-size} shows the ratio $\ell[enc({\cal T})]/\ell[\reprc]$ for Project Gutenberg books. It is seen that the codebook length is relevant both for words and syllables even for fairly long texts. It is  quantitatively less relevant for syllables than for words, because a given text has less different syllables than different words; hence (\ref{nebed4}) is smaller for syllables. Also, the codebook length is less relevant within $L={\rm variable}$ representation as compared to $L=8$, since the former codebook representation is more compact; see Fig.~\ref{fig:codebook-vs-code-size}.

The codebook length is irrelevant for the alphabet of letters. Indeed, for the alphabet of 26 English letters, the estimation of (\ref{eq:codebook-length-block-compr}) is straightforward; cf.~(\ref{nebed5}). We get that the codebook length is a modest number $\lesssim 200$, which for an average book is much smaller than other involved bit-lengths.

Let us provide some estimates for (\ref{entropy}) and (\ref{eq:codebook-length-block-compr}) for the alphabet of words. Recall that the ranked word frequencies $f_k$ in a typical English text approximately hold Zipf's law \cite{shannon2,grig}:
\BEA
\label{zi}
f_k= 1/(k\,C), \qquad C=\sum_{k=1}^{\m_{\rm words}}\frac{1}{k}\simeq \ln \m_{\rm words}, 
\EEA
where $C$ is the normalization, and where $\m_{\rm words}$ is the number of distinct words in the text. $N_{\rm words}$ is the total number of words in the text. Zipf's law is not valid for the whole range of frequencies \cite{pre}, but can be still employed for rough estimates \cite{shannon2,grig}.

\comment{In particular, (\ref{zi}) can be employed for estimating parameter $t$ in (\ref{eq:codebook-in-blocks}, \ref{nebed1}): $t\sim \lceil \log_2\frac{1}{f_1}\rceil$. Parameter $T$ in (\ref{eq:codebook-in-blocks}, \ref{nebed1}) is estimated from lowest frequency $\frac{1}{N_{\rm words}}$, since there are many words in texts that appear only once ({\it hapax legomena}, for which (\ref{zi}) does not hold) \cite{pre}. Hence, $T\sim\log_2N_{\rm words}$.}

Now in (\ref{nebed1}) we can assume that each word has in average 4.7 letters, take $\ell[code(a_k)]\simeq -\log_2 f_k\simeq \log_2(k\ln \m_{\rm words})$ and neglect $c$ in (\ref{or}). These lead from (\ref{entropy}) 
\BEA
\label{grund1}
&& \ell[\reprc]\gtrsim 4.7\times L\m_{\rm words},\\
&& \ell[enc({\cal T})]\simeq N_{\rm words}\log_2\left(\sqrt{\m_{\rm words}}\ln \m_{\rm words} \right).
\label{grund2}
\EEA
Fig~\ref{fig:word-count-histogram} shows the distribution of word counts in Project Gutenberg. For a text with typical values $\m_{\rm words}\sim 10^4$, $N_{\rm words}\sim 1.8\times 10^5$, and $L=8$, we get from (\ref{grund1}, \ref{grund2}): 
$\ell[enc({\cal T})]/\ell[\reprc]< 4.72$. This is consistent with the results on Project Gutenberg texts; see Fig.~\ref{fig:codebook-vs-code-size}. 

Eqs.~(\ref{grund1}, \ref{grund2}) show that there are two ways to make the codebook length small as compared to the code-only length for words: to take $\m_{\rm words}$ small, or making $N_{\rm words}$ large (for a fixed $\m_{\rm words}$). Books written by humans for humans do choose none of these ways, possibly because a large $N_{\rm words}$ is not manageable for reading, while a small $\m_{\rm words}$ is not interesting for reading. 

\subsubsection{Optimal alphabets}

As expected from Appendix \ref{when}, we found that more concatenated alphabets|e.g. letter 2-grams compared to letters or words compared to syllables|provide a smaller code-only length $\ell[enc({\cal T})]$; see Fig.~\ref{fig:code-only-length} that contrasts words with syllables. It is seen that in both cases $\ell[enc({\cal T})]$ increases with the number of words in the text. 

Fig.~\ref{fig:total-code-length} implies that the situation changes when we consider the total code-length $\ell[enc({\cal T})]+\ell[\reprc]$. Let us first assume that the codebook length is included within $L=8$ representation; see section \ref{godoy} and (\ref{nebed4}). Now for $81\%$ of Project Gutenberg texts syllables provide a lower total code length than for words, as Table~\ref{table1} shows. It also shows that for $\simeq 50\%$ of these texts syllables provide the best compression alphabet as compared to letters, 2-grams of letters and words. Hence syllables is the best alphabet for $L=8$.

The situation changes again within the more compact $L={\rm variable}$ codebook representation, which focuses on 26 letters and uses for them Huffman codes constructed from fixed letter frequencies; cf.~section \ref{godoy}. 
Table~\ref{table1} shows that words provide the best compression (smallest total code-length) over all alphabets for $81.6\%$ of texts. This is because within $L={\rm variable}$ the weight of the code only length $\ell[enc({\cal T})]$ in the total code-length $\ell[enc({\cal T})]+\ell[\reprc]$ is larger. Recall from Appendix \ref{when} that alphabets with longer (more concatenated) symbols have a competitive advantage.

\begin{figure}[!h]
\centering
    \subfigure[]{
    \includegraphics[width=8.6cm]{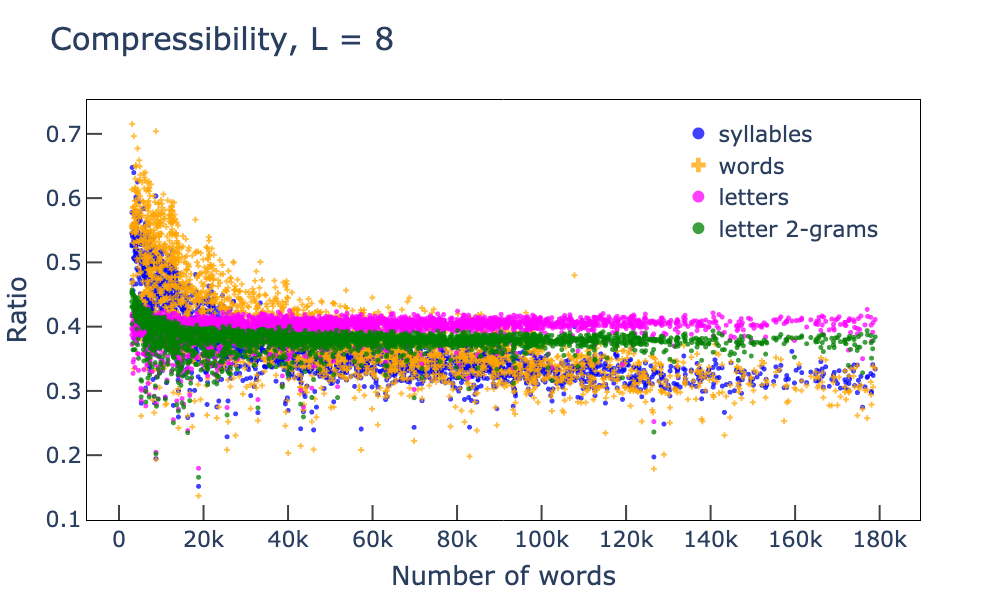}
    \label{fig:compressibility_L=8} 
    }
    \subfigure[]{
    \includegraphics[width=8.6cm]{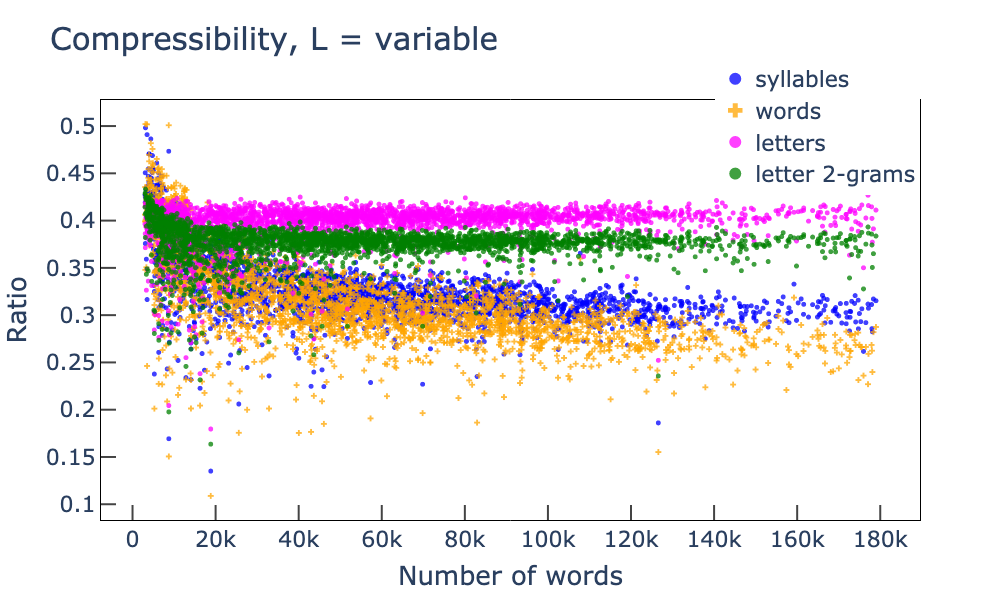}
    \label{fig:compressibility_L=variable} 
    }
    \caption{
    \comment{ (a)The normalized total code-length, i.e. $\frac{1}{N_{\rm words}}(\ell[enc({\cal T})]+\ell[\reprc])$ versus $N_{\rm words}$, where $N_{\rm words}$ is the total number of words in the text. The parametrization of the codebook length $\ell[\reprc]$ is done via $L=8$; cf.~section \ref{godoy}. Blue, orange, magenta and green points refer (resp.) to syllables, words, letters and letter 2-grams; cf.~Fig.~\ref{fig:code-only-length}. For short texts, $N_{\rm words}<10{\rm k}=10^4$, the alphabet of letters wins, i.e. it provides the lowest value of the total code-length. For $10 {\rm k}<N_{\rm words}\lesssim 50 {\rm k}$ the pairs of letters are better than letters and syllables. For $N_{\rm words}>50 {\rm k}$ (i.e. for sufficiently long texts) syllables win. 
    }
    (a) Compressibility $\eta$ defined via (\ref{fox}) versus $N_{\rm words}$, where $N_{\rm words}$ is the total number of words in the text. Codebook letters are represented via $L=8$; cf.~section \ref{godoy}. Blue, orange, magenta and green points refer (resp.) to syllables, words, letters and letter 2-grams. It is seen that
    all methods compress, since $\eta<1$. However, for short texts the letters provide smaller $\eta$, for moderate $N_{\rm words}$ the pairs of letters a give smaller $\eta$, while for long texts syllables win.\\
    (b) The same as in (a) but with representation $L={\rm variable}$ of codebook letters; cf.~section \ref{godoy} and (\ref{nebed5}). Words win in compressibility for $N_{\rm words}\geq 60$ k. 
    }
\end{figure}

These regularities are seen for the compressibility $\eta$ defined via (\ref{fox}); cf.~Figs.~\ref{fig:compressibility_L=8} and \ref{fig:compressibility_L=variable}. All methods compress, i.e. $\eta<1$, though short texts are not well compressed: $0.5<\eta<1$. Longer texts are compressed better by syllables or words; now $\eta\gtrsim 0.3$. For letters and letter 2-grams $\eta$ is larger and is nearly constant for not short texts, as Figs.~\ref{fig:compressibility_L=8} and \ref{fig:compressibility_L=variable} demonstrate.

We emphasize that our results depend on the codebook representation (\ref{eq:codebook-in-blocks}, \ref{dodo}), whose length enters into the total code length. It is then interesting to compare results produced via (\ref{eq:codebook-in-blocks}, \ref{dodo}, \ref{nebed1}) with those based on a uncompressed (naive) representation of (\ref{codebook}) presented in Appendix \ref{univ}. Such representations are necessarily longer than (\ref{nebed1}), hence they will provide a relative advantage to alphabets that have smaller number of symbols $M$. Hence, within the uncompressed representation of codebooks, syllables will have advantage with respect to words at least for not very long texts. For long texts the advantage is counter-balanced by the code only length contribution into the total code length. 

\begin{figure}[!h]
\centering
    \subfigure[]{
    \includegraphics[width=8.6cm]{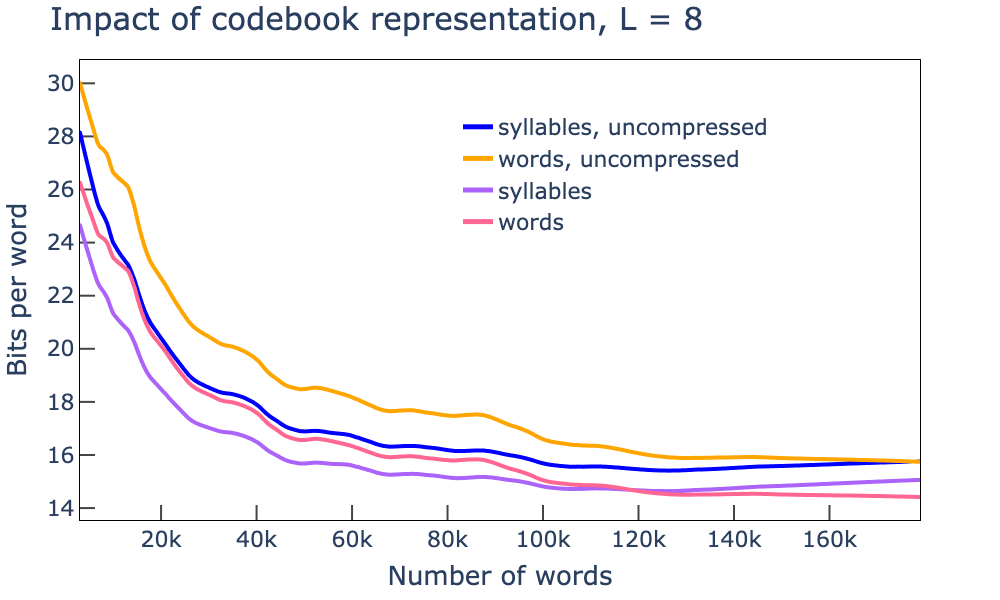}
    \label{fig:adi1} 
    }
    \subfigure[]{
    \includegraphics[width=8.6cm]{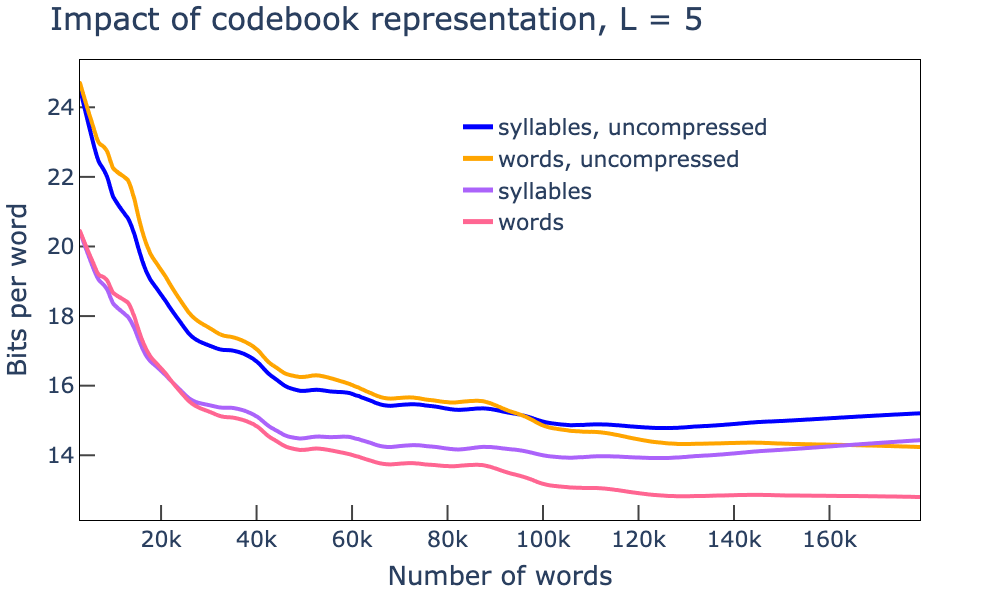}
    \label{fig:adi2} 
    }
    \caption{The normalized total code-length, i.e. $\frac{1}{N_{\rm words}}(\ell[enc({\cal T})]+\ell[\reprc])$ versus $N_{\rm words}$ for alphabets of words and letters, where $N_{\rm words}$ ("Number of words") is the total number of words in the text. Smoothing and interpolation were performed on the data. The parametrization of the codebook length $\ell[\reprc]$ is done via $L=8$ (a) or via $L=5$ (b); cf.~section \ref{godoy}. Two alternative codebook representations are compared: (\ref{eq:codebook-in-blocks}, \ref{dodo}) that is fit to Huffman coding {\it versus} the uncompressed representation from Appendix \ref{univ}. 
    }
\end{figure}

These effects are indeed seen in Figs.~\ref{fig:adi1} and \ref{fig:adi2} which compare the alphabets of words and syllables for texts of various lengths and for two representations: the compact representation (\ref{eq:codebook-in-blocks}, \ref{dodo}) {\it versus} the uncompressed representation (\ref{huk1}). Also, comparing with each other Fig.~\ref{fig:adi1} and Fig.~\ref{fig:adi2} we confirm that larger values of $L$ in (\ref{nebed4}) prioritize alphabets with smaller number of symbols $M$; i.e. syllables versus words.

\begin{figure}[!h]
\centering
 \subfigure[]{
    \includegraphics[width=8.6cm]{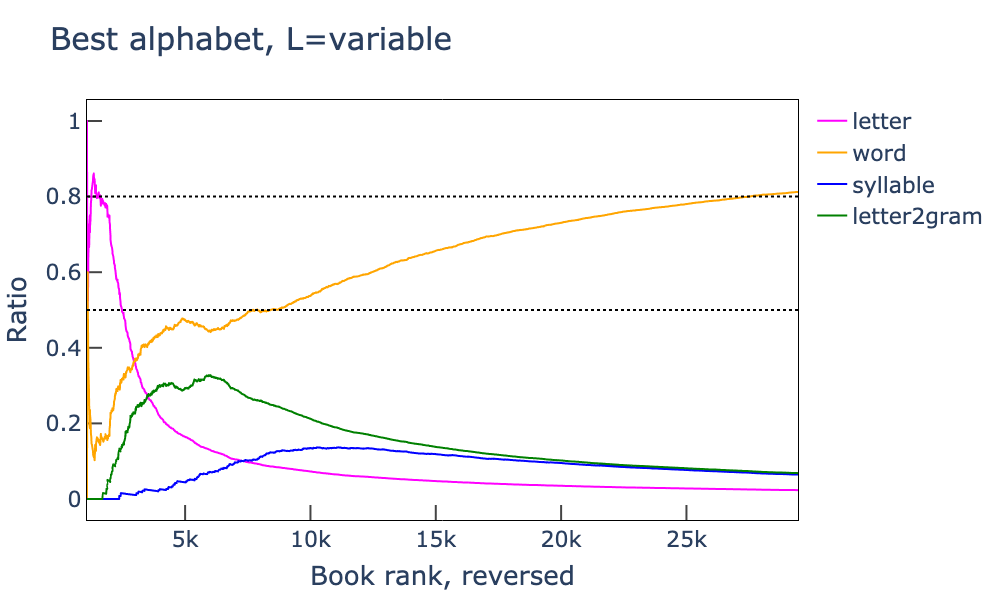}
    \label{fig:4a} 
    }
    \subfigure[]{
    \includegraphics[width=8.6cm]{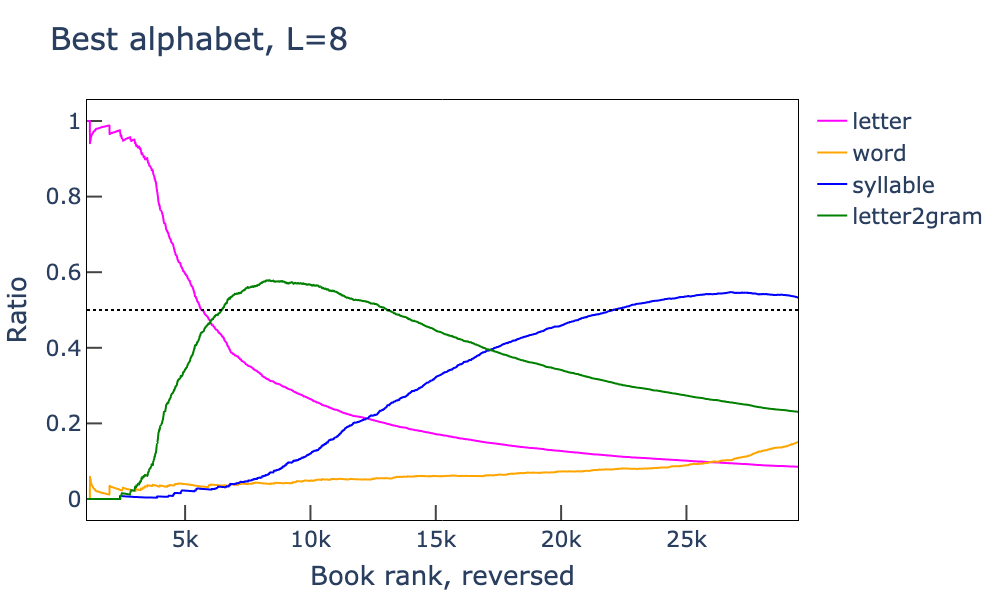}
    \label{fig:4b} 
    }
    \subfigure[]{
    \includegraphics[width=8.6cm]{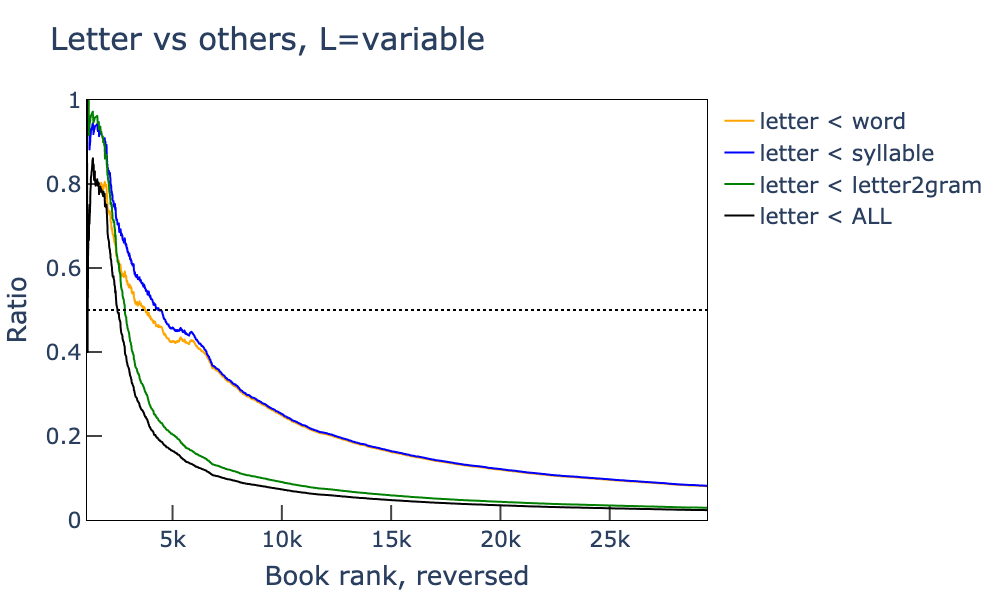}
    \label{fig:4c} 
    }
    \subfigure[]{
    \includegraphics[width=8.6cm]{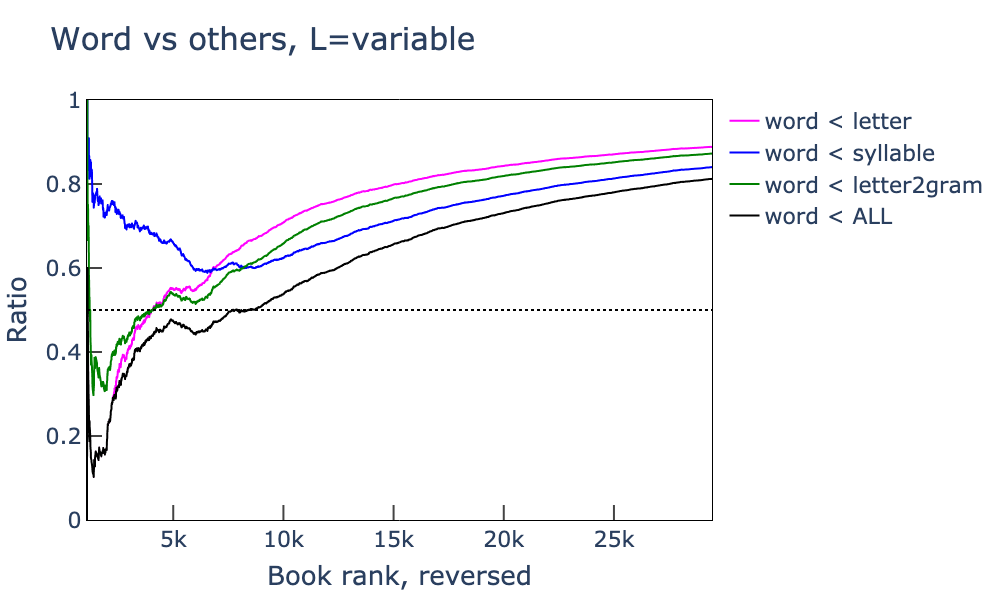}
    \label{fig:4d} 
    }
    \caption{Comparison between various alphabets for the 10\%-random sample of all 29708 Project Gutenberg books. After sampling all books retained their original ranks, curves were iterpolated. \\
    The $x$-axes of these figures shows the rank of all Project Gutenberg books: shorter books (the length is measured in words) books got smaller ranks. The $y$-axes shows various ratios (percentages): $y(x)$ is the corresponding percentage calculated for all books (of the sample) with rank $\leq x$. Everywhere the optimality statements (i.e. best or better) is understood in terms of the total compression length $\ell[enc({\cal T})]+\ell[\reprc]$ [see (\ref{nebed1})] provided by a concrete alphabet: syllables, words, letters and letter 2-grams. $L=8$ and $L={\rm variable}$ refer to different codebook representations; see section \ref{godoy}. \\ 
    (a) shows that for $L={\rm variable}$ the best compression alphabet is|depending on the text length|letters and words.\\ 
    (b) shows for $L=8$ that the best compression alphabet is letters, 2-grams of letters and syllables. \\
    (c) and (d) compare (resp.) alphabets of letters and words with other alphabets. Here "${\rm letter<word}$" means the ratio of all books where (till that rank), where letters fare worse than words, while "${\rm letter<ALL}$" means the ratio of books, where letters are the worst among words, syllables, and letter 2-grams. }
    \label{fig:4}
\end{figure}

\subsubsection{Optimal alphabets for texts filtered over their lengths}

It is interesting to examine the optimal alphabet when texts are limited by their length (in words). Hence,
Figs.~\ref{fig:4a}--\ref{fig:4d} show detailed competition results for the minimal total code-length. 
The $x$-axes of these figures shows the rank of all Project Gutenberg books: shorter books (the length is measured in words) books got smaller ranks. The $y$-axes shows various ratios (percentages): $y(x)$ is the corresponding ratio calculated for all books with rank $\leq x$.  

Figs.~\ref{fig:4a}, \ref{fig:4b} and \ref{fig:4c} demonstrate that for very short books (texts) the optimal alphabet is that provided by letters; i.e. for the sure majority of short texts the total code-length is smaller for the letter alphabet. Moreover, letters compete with letter 2-grams only, since the curve where letters are better than 2-grams coincides with the curve where the letters are better than all other alphabets. As expected, for $L=8$ the advantage of the letter alphabet extends for longer texts than for $L={\rm variable}$.

For longer texts the optimal alphabet essentially depends on the representation of letters in the codebook, i.e. on $L=8$ {\it versus} $L={\rm variable}$. Therefore, these cases should be considered separately. 

For $L=8$ the advantage goes to 2-grams for texts with $6{\rm k}\lesssim {\rm rank}\lesssim 17 {\rm k}$; see Fig.~\ref{fig:4b}. There is a range of books, where 2-grams do provide the optimal alphabet for a majority of books. For longer texts the advantage goes to syllables, which are optimal for nearly 50\% of all texts; see Table~\ref{table1}. Hence, syllables is the best alphabet for $L=8$. The sub-optimal alphabet for all texts is that of 2-grams of letters; see Fig.~\ref{fig:4b} and Table~\ref{table1}. 

In contrast, for $L={\rm variable}$ words take over the letters, and stay the optimal alphabet for all ranks larger than 5 k; see Figs.~\ref{fig:4a} and \ref{fig:4d}. Moreover, words provide the best alphabet for a majority of texts for ranks larger than 10 k; see Table~\ref{table1}. Note that pairs of words are not discussed, since they are always worse (with respect to the total code-length) than other alphabets.

\subsubsection{Letter $\n$-grams }

\begin{figure}[!h]
\centering
    \subfigure[]{
    \includegraphics[width=8.6cm]
    {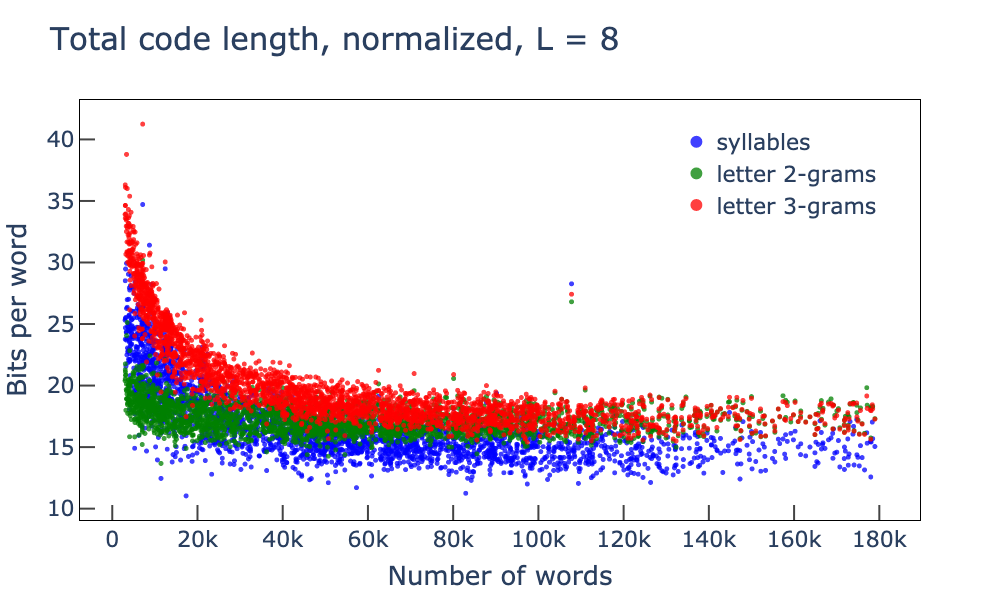}
    \label{fig:77}
    }
    \subfigure[]{
    \includegraphics[width=8.6cm]
    {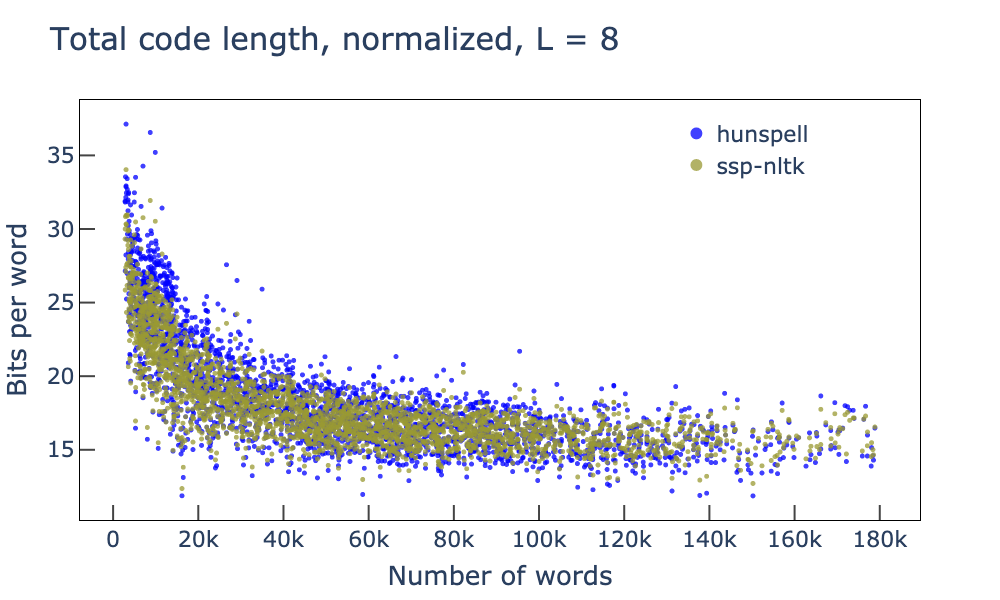}
    \label{compar} 
    }
    \caption{(a) For Gutenberg project books we show|for alphabets of syllables, letter $3$-grams and letter $4$-grams|the normalized total code-length  $\frac{1}{N_{\rm words}}\left(\ell[enc({\cal T})]+\ell[\reprc]\right)$ (denoted by "Bits per word"). Here $N_{\rm words}$ is the total number of words in each text; the codebook length $\ell[\reprc]$ is calculated according to $L=8$; cf.~section \ref{godoy} and (\ref{nebed1}, \ref{nebed4}). For sufficiently long texts syllables are better than 2-grams, while 2-grams are always better than 3-grams. \\
    (b) The same as in (a). Blue and green points refer to syllabification libraries implemented in (resp.) \cite{hunspell} and \cite{nltk}. They are denoted as "hunspell" and "ssp-nltk", respectively. It is seen that both libraries produce comparable results.
    }
\end{figure}

Letter 3-grams (and 4-grams) always provide a larger total code-length as compared to 2-grams; cf.~Fig.~\ref{fig:77}. One reason for this is that the number of distinct 3-grams is large ($\leq 26^3$). Hence their contribution to the codebook length is bigger than for 2-grams. 

Now compare these results with two facts. First, recall that the average length of an English syllable is $\simeq 3$ letters. Therefore, the syllables cannot be replaced by $\n$-grams of approximately the same length. Second, recall the known result by Shannon and others \cite{shannon2,dembo} that if only the length of the coded text is retained (i.e. the code-book length is unjustly omitted), the length of the optimal $\n$-gram is $\n\simeq 15$. It is seen that the real optimal $\n$-gram is really smaller than this $\n\simeq 15$ bound.

\section{Summary and discussion}
\label{summary}

\subsection{The problem and its solution}

We wanted to determine the best alphabet for the optimal (Huffman's) text encoding and compression. We think that solving this problem systematically will reveal important features of texts written by humans for humans, and will eventually improve practical schemes of text compression. 
The freedom of choosing the alphabet means that every text can be considered as a string of letters, $\n$-grams of letters, syllables, words, pairs of words {\it etc}. Despite the attention devoted to text compression within information theory, the optimal alphabet was so far not studied systematically. Solving this problem for a single text is not straightforward, since the compression involves the codebook that is text-specific and non-negligible for a typical text; see section \ref{zipo} and Fig.~\ref{fig:codebook-vs-code-size}. Hence the codebook should be regarded as an integral part of the compressed text, leading us to define the total code-length as the compressed text (code-only) length plus the codebook length, both expressed in bits; see section \ref{co}. An important property of the codebook is that its encoding scheme should be independent from the text, so that the decoder can know it in advance. Otherwise, we would need a (text-agnostic) codebook for codebook, which does not seem to us a useful concept at this point. Hence, we represented the codebook via bits in a compact and self-delimiting way; cf.~section \ref{codo} and \cite{if3}. 
\comment{
A text-specific compression of the codebook is not studied in this work, since it involves several new factors: we need to choose anew the alphabet for the codebook compression, design a text-agnostic codebook for codebook, understand the influence of hapax legomena (i.e. rare words) {\it etc}. }

Without accounting for the codebook length the solution to the problem of optimal alphabet selection is straightforward, since it amounts to using possibly more concatenated alphabets; see Appendix \ref{when}. In particular, one can use to this end sufficiently long blocks of letters ($\n$-grams) \cite{shannon2}. Entropy estimations \cite{shannon2} and numerical results \cite{dembo} show that $\n\simeq 15$ suffices for English texts.
If the codebook length is taken into consideration, the situation is drastically altered even for letter $n$-grams, since now $\n\leq 2$ for optimal $\n$-grams;  cf.~section \ref{results}. 

For texts of Project Gutenberg we found that the notion of the optimal alphabet depends on the length of texts and on the codebook representation; cf.~section \ref{godoy}. We focused on two such representations. One of them (denoted by $L=8$) employs the standard representation of alphabet symbols via ASCII codes of their letters. The second one ($L={\rm variable}$) is more compact, nearly the most compact within the letter representation of the codebook alphabet symbols; cf.~section \ref{godoy}.

For the $L=8$ codebook representation, the optimal alphabet is|depending on the increasing text length|letters, 2-grams of letters, and syllables; cf.~Figs.~\ref{fig:4b} and Table~\ref{table1}. No compression advantage is provided by words, pairs of words or $\n$-grams of letters with $\n>2$. The optimality of letters for short texts is natural, since for letters the burden of the codebook length is negligible in contrast to other alphabets. 
For the $L={\rm variable}$ representation, the letter alphabet is optimal for short texts, but for longer text the advantage goes to words which are optimal for $\simeq 81\%$ of Gutengerg Project texts; cf.~Figs.~\ref{fig:4a} and Table~\ref{table1}. 

Thus, for sufficiently long texts the optimal alphabet relates to a text representation via meaningful elements: syllables or words of the text. Pertinent differences between syllables and words are summarized as follows. {\it (i)} Syllables are less numerous than words: English has about 12000 syllables versus $\simeq 0.5\times 10^6$ words \cite{levelt}. Hence, the syllabic alphabet tends to be more compact also for a single text. {\it (ii)} Syllables are shorter than words. {\it (iii)} Not all syllables are meaningful (in contrast to words), but their structure is far from random and reflects the language evolution \cite{kessler,levelt}. Indeed, this is seen also in our results: each English syllable contains in average three letters, but 3-grams produce a much worst compression alphabet than syllables. As a result of {\it (i)} and {\it (ii)}, the alphabet of syllables has a shorter codebook, but a longer code only length (due to {\it (ii)}). These factors contribute into the interplay between words and syllables for texts of various length; see Figs.~\ref{fig:adi1} and \ref{fig:adi2}.

\subsection{Outlook}

In the context of our results, we make several remarks that could frame future studied.

Compact and self-delimiting codebook representations (\ref{eq:codebook-in-blocks}, \ref{dodo}) allowed us to consider two pertinent tasks; see section \ref{co}. First, we defined compressibility, i.e. we can meaningfully determine how much a given text is compressed compared to its original length. Second, we suggested a bound for the Kolmogorov complexity of a text; see Appendix \ref{kolmo}. Here there is a room for analytical calculations that we did not explore so far. It relates to feasible estimates of the codebook length in (\ref{nebed1}), which allows comparing Kolmogrov complexity to Lempel-Ziv complexity, a popular complexity measure with numerous interdisciplinary applications \cite{lz1,lz2}. 
In general, several points need to be clarified between the Lempel-Ziv data compression method and the approach described here.

The leitmotif of our work is that more concatenated alphabets provide better compression when the codebook length is not accounted for. But once this is done there is an optimal degree of concatenation.  
This situation is similar to probabilistic inference, where one can choose multi-parametric models to fit given data. Usually, increasing the number of parameters results in a better fit, eventually turning to overfitting. But when the complexity of the model is accounted for, the optimal number of parameters is finite, thereby preventing overfitting; see \cite{grunwald,mathpsy} for a review. 

We can gain an interesting perspective on our results by recalling that every human language has at least two structures \cite{boer}: The {\it combinatorial} structure (or dual patterning) means that meaningless elements of lower hierarchical levels (letters or phonemes) combine to form meaning-expressing morphemes and words. 
Syllables are somewhere in between, since there are both meaningful syllables (e.g. monosyllabic words, but not only them) and meaningless syllables; e.g. just a phoneme. 
Hockett proposed the dual patterning as a design feature of human languages and also offered a hypothesis for its emergence: it developed for keeping the meaning-expressing elements sufficiently different \cite{hockett,nowak}. The {\it compositional} structure refers to the fact that a finite number of meaningful words make up phrases and sentences expressing a huge variety of meanings \cite{boer}. 
The two structures overlap, e.g. because the phonemes can participate in the meaning-formation, as witnessed in sound symbolism \cite{sound}. Moreover, phonemes participate in text-formation not only via words, but also directly \cite{deng}.

Hence, despite the fact that we employ statistical data compression that does not account directly for the meaning of studied texts, our results uncover a new dimension in the hierarchical (combinatorial-compositional) structure of the language: the compression of a sufficiently long text is maximal when one structure changes for another.

\acknowledgements This work was supported by SCS of Armenia, grant No. 21AG-1C038.

\appendix

\section{Syllabication: short review and examples}
\label{ap_a}

Syllable is a unit of pronunciation which consists (in the linear order) of onset, nucleus and coda. Onset and coda are optional, the nucleus contains vowel(s) and is obligatory \cite{sound}. 
Statistically, the nucleus is more closely related to its coda, then to the onset \cite{kessler}. With different variations, these features of syllable hold across different languages; e.g. in Mandarin Chinese, where there are 23 onsets (mostly single consonants), and the (semantically important) tone of the syllable relates to the nucleus+coda \cite{chin}. 

Precise definitions of syllable are technically complex \cite{kornai}. Rules of syllabification have exclusions, and sometimes contradict each other \cite{kessler,frau}. Hence, syllabification practices have to rely on conventions; see \cite{hunspell} and \cite{nltk} for two widespread syllabication systems. Nevertheless, it is useful to discuss briefly the available syllabification rules, also because the advantage of syllables as compression alphabets can lead to creating a new syllabication system that will be purposefully tuned to needs of text compression. 

{\it (1)} Do not divide one phoneme between two syllables. In particular, diphthongs \footnote{There are eight English diphthongs:
a\i\, (m\underline{y}), a\textupsilon\, (h\underline{ow}),
o\textupsilon\, (g\underline{o}), e\i\, (d\underline{ay}), \i\textschwa\, (h\underline{ere}), o\i \,
(b\underline{oy}), \textupsilon \textschwa\, (t\underline{our},
p\underline{ur}e), e\textschwa \, (w\underline{ear}, f\underline{air}).}, and mute vowels are not counted as nuclei of syllabification. 

{\it (2)} Keep morphemes intact, even when this contradicts the pronunciation; e.g. prefer {\it con-stant} to {\it cons-tant}, and {\it car-oli-na} to {\it ca-ro-li-na}.

{\it (3)} Divide between two middle consonants. For example:
{\it hap-pen, let-ter, din-ner}. But do not contradict to {\it (1)}, i.e. do not divide {\it th, sh, ph, th, ch, wh}. Also, do not divide double consonants {\it ss} and {\it ll}, whenever it contradicts {\it (2)}: {\it bless-ing}, {\it pass-word}, {\it fall-en}. 

{\it (4)} Legality principle \cite{frau}: prefer onsets that can be a beginning of a word, and codas that can be an end of word; e.g. {\it ad-mit} and not {\it adm-it}.

\comment{
{\it (5)} The maximal onset principle \cite{frau}: prefer longer onsets possibly without contradicting to {\it (1)} and to {\it (4)}.}

{\it (5)} Sonority Sequencing Principle (SSP) starts with distributing all phonemes over the sonority scale \cite{frau,ssp,zec} \footnote{Here we mention how some of phonemes are distributed in this scale (from higher sonority to lower, [e,o] means that phonemes [e] and [o] have the same sonority): [a], [e, o], [i, u, j, w], [l], [m, n], [z, v], [f, $\theta$, s], [b, d, g], [p, t, k].}. Now SSP states that moving from onset to  nucleus relates to the sonority increase, which is peaked at the nucleus, because moving from the nucleus to coda implies a sonority decrease. SSP explains why English syllables {\it matl} and {\it lkon} are impossible: in {\it matl} the sonority in the sequence {\it tl} increases (must be decreasing according to SSP) and in {\it lkon} the sonority of the sequence {\it lk} decreases (must be increasing). SSP does have its limitations among English syllables \cite{zec}. For example, the syllable-word {\it trust} holds SSP; while {\it Spain} violates it. However, a viable syllabification system NLTK is based on SSP \cite{nltk}.

Since different syllabification systems are available, it is natural to ask how they compare with each other as compression alphabets. Fig.~\ref{compar} answers this question for two widespread syllabification methods: hunspell and ssp-nltk described in (resp.) \cite{hunspell} and \cite{nltk}. Now ssp-nltk more frequently applies the SSP principle [see {\it (5)} above], while hunspel applies other rules of syllabification. Recall that hunspell is employed in the main text as the main syllabification method. 

Fig.~\ref{compar} shows that from the viewpoint of compressing the total code-length 
$\ell[enc({\cal T})]+\ell[\reprc]$ the two methods are nearly equivalent, though ssp-nltk (not employed in the main text) has an advantage: it does somewhat smaller total code-length when averaged over all texts of Project Gutenberg. Hence using ssp-nltk instead of hunspell will only improve the performance of the syllabic alphabets as compared to other alphabets. 

\section{An uncompressed codebook representation}
\label{univ}

The code representation (\ref{eq:codebook-in-blocks}, \ref{dodo}) discussed in section \ref{codo} applies for Huffman and Shannon codes. It does not apply to an arbitrary coodebook defined in (\ref{codebook}). 
We now describe a decodable (but uncompressed) binary representation of the codebook ${\cal C}(\cal T, \cal A)$ that applies to all codebooks:
\BEA
\hreprc=\halpha(a_1)\gamma(code(a_1))\;\halpha(a_2)\gamma(code(a_2))\;....\; \halpha(a_n)\gamma(code(a_n)),
\label{huk1}
\EEA
where $\halpha(a_k$) is a representation of an alphabet symbol $a_k$ and where $\gamma(\cdot)$ is defined in (\ref{eq:gamma-code}). It encodes each letter of $a_k$ into a bit-sequence of length $L+1$, where the first bit is a "marker" bit and is always 1, followed by $L$ bits that encode the letter. This $L$-bit letter-to-bits mapping is shared beforehand between the encoder and the decoder. For instance, $L=8$ if we decide to employ the standard ASCII code for letters, or $L=5$ if we lemmatize our texts by keeping the 26 English letters plus a few (not more than 32-26=6) additional text symbols. 

To decode $\hreprc$, the decoder reads a 1 and then the next $L$ bits to decode a letter, and repeats this until it sees a $0$ at the "marker" bit position -- this means it has reached the start of $\gamma(code(a_1))$.  It then decodes $\gamma(code(a_1))$ to obtain $code(a_1)$. As $\gamma(\cdot)$ is self-delimiting, the decoder knows where it ends and where the next "line" of the codebook starts. Hence the bit representation can be decoded uniquely without text-specific prior knowledge, and the bit-length $\ell[\hreprc]$ of $\hreprc$ reads
\BEA
\ell[\hreprc]=(L+1)\times letters({\cal A})+\sum_{k=1}^\m{\ell[code(a_k)]} + 2\sum_{k=1}^{\m}{\log_2(\ell[code(a_k)])},
\label{huk2}
\EEA
where the last term in (\ref{huk2}) is normally small compared to other two terms.

\section{Codebook encoding via letter}
\label{codebook_letters}

Frequencies of English letters taken from \cite{wiki_letters}: 'a': 0.082, 'b': 0.015, 'c': 0.028, 'd': 0.043, 'e': 0.13,'f': 0.022, 'g': 0.02, 'h': 0.061, 
  'i': 0.07, 'j': 0.0015, 'k': 0.0077, 'l': 0.04,
  'm': 0.024, 'n': 0.067, 'o': 0.075, 'p': 0.019,
  'q': 0.00095, 'r': 0.06, 's': 0.063, 't': 0.091,  
  'u': 0.0028, 'v': .0098, 'w': 0.024, 'x': 0.0015,  
  'y': 0.02, 'z': 0.0074. For convenience, these frequencies are not normalized, their sum slightly exceeds $1$. 

According to Huffman's method, letters got the following codewords:
'a': '1110',
'b': '110000',
'c': '01001',
'd': '11111',
'e': '100',
'f': '00100',
'g': '111100',
'h': '0110',
'i': '1011',
'j': '001011011',
'k': '0010111',
'l': '11001',
'm': '00110',
'n': '1010',
'o': '1101',
'p': '110001',
'q': '001011000',
'r': '0101',
's': '0111',
't': '000',
'u': '01000',
'v': '001010',
'w': '00111',
'x': '001011010',
'y': '111101',
'z': '0010110011'

\section{Concatenated alphabets provide lower code-only length}
\label{when}

\subsection{Statement of the result}

Consider two alphabets ${\cal W}=\{w_k\}_{k=1}^{n_w}$ and ${{\cal S}}=\{{s}_k\}_{k=1}^{n_s}$, such that each symbol $w_k$ is a concatenation of one or more symbols from ${\cal S}$; e.g. ${\cal W}$ and ${\cal S}$ can be (resp.) distinct words and distinct syllables of a given text ${{\cal T}}$. Without loss of generality we continue the discussion in terms of this example. Now words from ${\cal W}$ consist of different number of syllables. Let the maximal number of syllables per word in ${{\cal T}}$ be $z$, while $\bar{z}$ is the average number of (real) syllables per word in ${{\cal T}}$. For English texts $\bar{z}\simeq 1.7$, while we can safely take $z=4-5$: larger values of $z$ are formally allowed, but are less useful (as seen below), moreover that such words are rare. 

Using features of entropy, section \ref{ap_aa} deduces the following relation between the entropy of words $S_{\rm words}$ in ${{\cal T}}$, the entropy of syllables $S_{\rm syllab}$,  $z$ and $\bar{z}$:
\BEA
\label{pu}
&&S_{\rm words}-zh_2[z/\bar{z}]\leq \bar{z}S_{\rm syllab},\\
&&h_2[x]\equiv-x\log_2[x]-(1-x)\log_2[1-x],
\label{h2}
\EEA
where $S_{\rm words}$ and $S_{\rm syllab}$ are defined via (\ref{entropy}) with (resp.) the frequencies of words and syllables in ${{\cal T}}$.

Let us now assume that we can neglect $zh_2[z/\bar{z}]$ in (\ref{pu}). This assumption hold for many real texts, as we checked. For example, it holds when $S_{\rm words}={\cal O}(\log_2\m_w)$ and/or $S_{\rm syllables}={\cal O}(\log_2\m_s)$, where the number of distinct words $\m_{\rm words}$ and/or the number of distinct syllables $\m_{\rm syllab}$ are sufficiently large. Then $zh_2[z/\bar{z}]\leq z={\cal O}(1)$ can be neglected in (\ref{pu}). Hence, after multiplying both sides of (\ref{pu}) by $N_{\rm words}$ we get:
\BEA
N_{\rm words}S_{\rm words}\leq N_{\rm syllab}S_{\rm syllab},
\label{pu2}
\EEA
where $N_{\rm syllab}$ ($N_{\rm syllab}$) is the overall number of syllables (words) in ${\cal T}$. Note that the assumption is not needed, and (\ref{pu2}) follows directly from (\ref{pu}), if $z=\bar{z}$. This is the case when comparing $\n$-grams of letters with $\widetilde{\n}$-grams ($\widetilde{\n}=p\n$, where $p$ is an integer), or pairs of words with words {\it etc}. Note that the literature tends to state results similar to (\ref{pu2}) in a limited form that is not suitable for our purposes, e.g. for a stationary random process \cite{cover}.

If the compressed text-length in (\ref{hu}, \ref{entropy}) is determined by the entropy (i.e. the influence of $c$ is not essential, as happens for many real texts), (\ref{pu2}) implies that the minimal length of $enc({\cal T})$ from (\ref{hu}, \ref{entropy}) decreases if we move to an alphabet with longer average symbols, e.g. go from words to syllables. Below we shall confirm this result for letters, $\n$-grams of letters ($\n=2,3,4$), syllables, words, and pairs of words. Each of these is a concatenated version of the previous alphabet. However, the total length of the code will not hold this relation: as seen in section \ref{results} there is an optimal degree of concatenation.

\subsection{Derivation of Eq.~(\ref{pu})}
\label{ap_aa}

Consider two alphabets ${\cal W}=\{w_k\}_{k=1}^{n_w}$ and ${{\cal S}}=\{{s}_k\}_{k=1}^{n_s}$, such that each symbol $w_k$ is a concatenation of one or more symbols from ${\cal S}$; e.g. ${\cal W}$ and ${\cal S}$ can be (resp.) distinct words and distinct syllables of a given text ${{\cal T}}$. Without loss of generality we continue the discussion in terms of this example. Now words from ${\cal W}$ consist of different number of syllables. Let the maximal number of syllables per word in ${{\cal T}}$ be $z$. Introduce an additional (empty) syllable $\Theta$. Adding $\Theta$ to end of words we can make every word to consist of the same number of syllables $z$. Note that adding $\Theta$ does not change the frequency of the word in ${{\cal T}}$. 

Let $f_{i_1...i_z}$ be the frequency of the word that consists of syllables $s_{i_1}..s_{i_z}$: $\sum_{i_1...i_z}f_{i_1...i_z}=1$. Define marginal frequencies for syllable $s_i$ to appear in position $u$:
\BEA
f_i^{[u]}=\sum_{i_1..i_{u-1}i_{u+1}..i_z}f_{i_1..i_{u-1}\,i\,i_{u+1}..i_z}, \qquad u=1,...,z.
\EEA
It should be clear that the frequency $g_i=g[a_i]$ of the syllable $s_i$ in ${\cal T}$ reads:
\BEA
g_i=\frac{1}{z}\sum_{u=1}^z f_i^{[u]}.
\label{str}
\EEA
Now employ first the sub-additivity \cite{cover} of entropy and then its concavity via (\ref{str}):
\BEA
\label{kru}
\frac{1}{z}S[f]&\equiv& -\frac{1}{z}\sum_{i_1...i_z}f_{i_1...i_z}\log_2 f_{i_1...i_z}\leq\frac{1}{z} \sum_{u=1}^z S[f^{[u]}] \\
&\leq& S[g]\equiv -\sum_ig_i\log_2 g_i.
\EEA
$S[g]$ is not yet the entropy $S_{\rm syllab}$ of real syllables, since $S[g]$ contains the probability of the auxiliary syllable $\Theta$. Put differently, $S_{\rm syllab}$ refers to frequencies of real (i.e. without $\Theta$) syllables only. Denote the frequency of $\Theta$ by $g_\Theta$ and note
\BEA
&&S[g]=h_2[g_\Theta]+[1-g_\Theta]S_{\rm syllab}, \\ &&h_2[g_\Theta]\equiv-g_\Theta\log_2[g_\Theta]-(1-g_\Theta)\log_2[1-g_\Theta].
\label{bru}
\EEA
Next, let us show that
\BEA
g_\Theta=1-\frac{\bar{z}}{z},
\label{ru}
\EEA
where $\bar{z}$ is the average number of real syllables per word in ${\cal T}$: $\bar{z}\equiv\sum_{k=1}^{n_w}f(w_k)s(w_k)$, where the sum goes over all elements of ${\cal W}$ ($f(w_k)$ is the frequency of the word $w_k$), and $s(w_k)$ is the number of real syllables in $w_k$. Indeed, the number of $\Theta$-syllables in ${\cal T}$ is $zN_{\rm words}g_\Theta$, where $N_{\rm words}$ is the overall number of words in ${\cal T}$. Now we have $zN_{\rm words}g_\Theta=N_{\rm words}\sum_{k=1}^{n_w}f(w_k)[z-s(w_k)]$, and (\ref{ru}) follows. 
Eqs.~(\ref{kru}--\ref{ru}) imply relation (\ref{pu}) for the entropy of words $S_{\rm words}=S[f]$.

\section{Kolmogorov complexity of texts}
\label{kolmo}

Kolmogorov complexity is frequently used (sometimes implicitly) in information theory and statistics \cite{cover,LiVitanyi}. Hence, we set to interpret above relations in terms of an upper bound for Kolmogorov complexity of a text. For any bit-string $x$, Kolmogorov complexity $K[x]$ is defined as the bit-length of the minimal program that {\it (i)} runs a universal computer (or Turing machine) from some standard state, {\it (ii)} prints $x$ and {\it (iii)} halts the computer \cite{LiVitanyi}. Note that $K[x]$ is defined with respect to a computer-dependent constant ${\cal O}(1)$, which is the length of the routine employed for translating from one universal computer to another \cite{LiVitanyi}. Below we omit this ${\cal O}(1)$ from formulas. 

For a given text ${\cal T}$ we now provide a program $P$ that holds the above conditions {\it (i)--(iii)}. $P$ can be represented as the following concatenated binary string [cf.~(\ref{hu}, \ref{eq:codebook-length-block-compr}, \ref{eq:gamma-code}, \ref{2star})]:
\BEA
\label{eq:Kolmogorov-tape-repr}
 P = r \, \gamma(enc({\cal T})), \quad r\equiv \reprc.
\EEA
To decode $P$, the decoder reads and decodes $r=\reprc$ -- this is possible because $r$ is self-delimiting, as described in section \ref{subsec:Codebook-CodebookRepresentation}. The decoder reads $enc({\cal T})$ and halts, because $\gamma(enc({\cal T}))$ is self-delimiting. 
Thus, we can bound from (\ref{eq:Kolmogorov-tape-repr}) the Kolmogorov complexity of a text [cf.~(\ref{2star})]
\BEA
\label{eq:Kolmogorov-complexity-text}
K[{\cal T}]\leq \ell[r]+ \ell[\gamma(enc({\cal T}))]=\ell[r]+\ell[enc({\cal T})]
+2\lceil\log_2 \ell[enc({\cal T})]\rceil,
\EEA
where $\ell[enc({\cal T}))]$ and $\ell[\reprc]=\ell[r]$ are recovered from (resp.) (\ref{or}, \ref{entropy}) and (\ref{nebed1}). In this context note that (\ref{nebed1}) can be expanded as follows using (\ref{2star}):
\BEA
\label{nebed2}
{\ell}[\reprc] &=& \sum_{a \in {\cal A}} \ell[\alpha(a)]+ \sum_{z=t}^T \left\{
\lceil \log_2k_z\rceil +\lceil \log_2z\rceil \right\}
\\
\label{nebed3}
&+&2\sum_{a \in {\cal A}} 
\lceil \log_2\ell[\alpha(a)]\rceil+ 2\sum_{z=t}^T \left\{
\lceil \log_2\log_2k_z\rceil +\lceil \log_2\log_2z\rceil \right\}+\ell[\gamma(T-t)],
\EEA
where we used $\lceil \log_2\lceil\log_2 z\rceil\rceil=\lceil \log_2\log_2 z\rceil$ with conventional definition $\log_2[x\leq 0]=0$. For sufficiently large texts, the major terms in ${\ell}[\reprc]$ are those given by (\ref{nebed2}).

\section{Alphabet encoding implementation}
\label{pseudo}

Algorithm \ref{alg:AlphabetEncoding} describes the procedure block-based encoding of the codebook presented in section \ref{codo}. The algorithm gets as input the text to encode, and functions \textit{tokenize} and \textit{letterEncode} (we use regular font for variables and italic for functions). \textit{tokenize} splits the text into tokens, as defined by the underlying symbol alphabet we are using. For example, if we want to use syllables as the symbols, we pass the appropriate \textit{tokenize} function which splits the text into syllables -- those become our tokens. \textit{letterEncode} encodes symbols into letters. Depending on whether we want to use the letter-encoder for $L=5$, $L=8$ or $L={\rm variable}$ (variable-length) codes, we pass the appropriate \textit{letterEncode} function.

\begin{algorithm}[htb!]
\label{alg:AlphabetEncoding}
\DontPrintSemicolon
\SetAlgoLined
\caption{Alphabet Encoding}\label{alg:AlphabetEncoding1}
\KwIn{text, $tokenize$, $letterEncode$}
\KwOut{tuple (code, codebook)}

tokens $\gets$ $tokenize$(text) \label{alg:line:tokenize}\\
code = huffCoder.$fitEncode$(tokens) \label{alg:line:huffmanFitEncode} \\
symbolsByLength $\gets \{\}$ \label{alg:line:wordsByLength-start} \\
\For{{\upshape s, c in huffCoder.encoding}} 
{\tcp{s is the symbol and c is its Huffman code} 
    symbolCode = $letterEncode$(s) \label{alg:line:letterEncode} \\
    symbolsByLength[$len$(c)].$append$(symbolCode) 
}\label{alg:line:wordsByLength-end}
codeBook $\gets$ $len$(symbolByLength) \tcp{$\gamma(T-t)$ in (\ref{nebed1})} \label{alg:line:block-encoding-start}
\For{\upshape{codeLen, symbolCodeList in symbolsByLength}} 
{
    codebook += $gamma$(codeLen)  \tcp{$\gamma(z)$ in (\ref{dodo})}
    codebook += $gamma$(len(symbolCodeList)) \tcp{$\gamma(k_z)$ in (\ref{dodo})}
    \For {\upshape{wordCode in symbolCodeList}}{
        codebook += $gamma$(symbolCode) \tcp{$\gamma(a_j^{(z)})$ (\ref{dodo})}
    }
} \label{alg:line:block-encoding-end}
return code, codebook

\end{algorithm}

The algorithm tokenizes the text at line \ref{alg:line:tokenize}, then obtains a Huffman code for those tokens on line \ref{alg:line:huffmanFitEncode}. On lines \ref{alg:line:wordsByLength-start}-\ref{alg:line:wordsByLength-end} we go through the Huffman encoding and obtain a letter-based code for each symbol using the function \textit{letterEncode}.
One line \ref{alg:line:letterEncode} we encode the symbol and obtain its letter-code. 
We add the letter-codes for all symbols which have the same code-length (in terms of the Huffman code at line \ref{alg:line:huffmanFitEncode}) into a list. We add those lists into a dictionary, where the key is the code-length, and the value is the list. Next, in lines \ref{alg:line:block-encoding-start} to \ref{alg:line:block-encoding-end} we encode the blocks of symbols as presented in (\ref{dodo}) and (\ref{nebed1}). The output of the function is a tuple consisting of the encoding of the text, and the codebook.

\end{document}